\pdfoutput=1
\documentclass[nofootinbib,letterpaper,floatfix,preprintnumbers,twocolumn]{revtex4}

\usepackage{color}

\usepackage{graphicx}
\usepackage{hyperref}
\usepackage[utf8]{inputenc}

\usepackage{lipsum}
\makeatletter
\def\ps@pprintTitle{%
 \let\@oddhead\@empty
 \let\@evenhead\@empty
 \def\@oddfoot{}%
 \let\@evenfoot\@oddfoot}
\makeatother

\usepackage{relsize}

\usepackage{verbatim}
\usepackage{amsmath}
\usepackage{amssymb}
\usepackage{subfigure}
\usepackage{url}

\usepackage{multirow}
\usepackage{threeparttable}
\usepackage{paralist}

\usepackage{color}
\definecolor{darkgreen}{rgb}{0,0.5,0}

\newcommand{\SU}{\text{SU}}
\newcommand{\SO}{\text{SO}}

\newcommand{\U}{\text{U}}

\DeclareRobustCommand{\Sec}[1]{Sec.~\ref{#1}}

\DeclareRobustCommand{\Tab}[1]{Table~\ref{#1}}

\DeclareRobustCommand{\Fig}[1]{Fig.~\ref{#1}}

\DeclareRobustCommand{\Eq}[1]{Eq.~(\ref{#1})}

\newcommand{\be}{\begin{equation}}
\newcommand{\ee}{\end{equation}}

\newcommand{\mb}[1]{\boldsymbol{#1}}

\newcommand{\hc}{{\rm h.c.}}
\newcommand{\ie}{{\it i.e.}}
\newcommand{\eg}{{\it e.g.}}
\newcommand{\nn}{\nonumber}
\newcommand{\order}{{\cal O}}

\begin{document}

\preprint{CERN-PH-TH-2015-022}

\title{Neutrino Masses from Neutral Top Partners}

\author{Brian Batell} 
\address{Theory Division, CERN, 1211 Geneva 23, Switzerland}

\author{Matthew McCullough} 
\address{Theory Division, CERN, 1211 Geneva 23, Switzerland}

\date{\today}

\begin{abstract}
We present theories of `Natural Neutrinos' in which neutral fermionic top partner fields are simultaneously the right-handed neutrinos (RHN), linking seemingly disparate aspects of the Standard Model structure:  a) The RHN top partners are responsible for the observed small neutrino masses, b) They help ameliorate the tuning in the weak scale and address the little hierarchy problem, and c) The factor of $3$ arising from $N_c$ in the top-loop Higgs mass corrections is countered by a factor $3$ from the number of vector-like generations of RHN. The RHN top partners may arise in pseudo-Nambu-Goldstone-Boson (pNGB) Higgs models such as the Twin Higgs, as well as more general Composite, Little, and Orbifold Higgs scenarios, and three simple example models are presented. This framework firmly predicts a TeV-scale seesaw, as the RHN masses are bounded to be below the TeV scale by naturalness. The generation of light neutrino masses relies on a collective breaking of lepton number, allowing for comparatively large neutrino Yukawa couplings and a rich associated phenomenology.
The structure of the neutrino mass mechanism realizes in certain limits the Inverse or Linear classes of seesaw.
Natural Neutrino models are testable at a variety of current and future experiments, particularly in tests of lepton universality, searches for lepton flavor violation, and precision electroweak and Higgs coupling measurements possible at high energy $e^+ e^-$ and hadron colliders.
\end{abstract}

\maketitle

\section{Introduction}
\label{sec:introduction}

The first run of the Large Hadron Collider has placed significant pressure on the hypothesis of natural electroweak symmetry breaking. 
This pressure comes from two sources. First, the measured properties of the Higgs boson are consistent with the predictions of the Standard Model (SM), whereas one would generically expect deviations from these predictions for a natural Higgs. Second, direct searches for new states responsible for softening the sensitivity of the Higgs mass to high scales, such as, \eg,  top partners, have so far turned up empty.  In particular, for the familiar case of colored top partners, such as scalar top squarks in supersymmetry or fermionic top partners in composite Higgs theories, strong constraints have resulted from these searches, in some cases pushing their allowed masses into unnatural territory~\cite{Craig:2013cxa,Matsedonskyi:2014mna,Dine:2015xga}. 

In light of this situation, there has been renewed interest in the idea of {\it neutral naturalness},  which hypothesizes that the partner states responsible for the cancellation of the quadratic divergences do not carry color charge. Such color-neutral partners are more difficult to constrain directly at the LHC due to their significantly smaller production cross sections, thus allowing more natural theories of electroweak symmetry breaking at the price of additional model complexity.  This novel approach to naturalness dates back to the Twin Higgs model~\cite{Chacko:2005pe}, and a number of interesting extensions and generalizations have been proposed in the literature~\cite{Barbieri:2005ri,Chacko:2005vw,Chacko:2005un,Chang:2006ra,Burdman:2006tz,Goh:2007dh,Poland:2008ev,Batra:2008jy,Cai:2008au,Craig:2013fga,Craig:2014aea,Geller:2014kta,Burdman:2014zta,Craig:2014roa,Craig:2015pha,Barbieri:2015lqa,Low:2015nqa}. 
Recent studies of the experimental constraints and signatures are presented in Refs.~\cite{Burdman:2014zta,Craig:2015pha}.  Color-neutral top partners are also increasingly motivated by the observed properties of the Higgs boson, which indicate that the Higgs-gluon-gluon and Higgs-photon-photon couplings are approximately SM-like, suggesting the Higgs may not be coupled to light colored or charged fields.

Intriguingly, there are other strong hints in nature for new neutral states beyond those present in the SM. Indeed, two of the most compelling empirical suggestions of new physics come from the need to generate neutrino masses and from the disparate gravitational phenomena pointing towards dark matter. Neutrino masses can be elegantly  explained by the introduction of new neutral fermions -- the right-handed neutrinos (RHN) -- which mix with the left-handed neutrinos via Yukawa interactions~\cite{Minkowski:1977sc,Yanagida,Glashow,GellMann:1980vs,Mohapatra:1979ia}.  Furthermore, the simplest dark matter candidates consist of new cosmologically stable neutral particles. A natural question to ask is whether these new neutral states, required to understand these empirical mysteries, can also play a role in naturalness. Stated more simply, can these neutral states be top partners? Remarkably, the potential connection between neutral top partners and dark matter has already been explored in an early paper by Poland and Thaler~\cite{Poland:2008ev}, who showed that neutral top partners could indeed serve as viable dark matter candidates. 

In this paper we focus on the previously unexplored possibility that the RHNs are simultaneously the top partners responsible for canceling the dominant quadratic divergence to the Higgs mass. Since the RHNs are fermions, we are naturally led to consider theories in which the Higgs is a pseudo-Nambu-Goldstone boson (pNGB)~\cite{Kaplan:1983fs,Kaplan:1983sm,Georgi:1984af,ArkaniHamed:2001nc,ArkaniHamed:2002qx,ArkaniHamed:2002qy,Contino:2003ve,Agashe:2004rs}, in analogy with the pions of QCD. Such theories allow for fermionic top partners, which are united with the top in a multiplet transforming under the spontaneously broken global symmetry of which the Higgs is a low energy remnant. In order for the top partners to be neutral under color, the $\SU(3)_c$ factor of the global symmetry must be enlarged to contain, in the most straightforward cases, an additional $\SU(3)$ factor, which is needed to account for the multiplicity factor in the top-partner loop.  Here we wish to speculate that this is in fact the flavor symmetry $\SU(3)_N$ which acts on the RHNs, thus making a tentative connection between the requirement of three top-partner fields for naturalness and the existence of three generations of RHN. Additionally, to enforce the cancellation of quadratic divergences, a $Z_2$ interchange symmetry or larger $\SU(6)$ symmetry (which contains $\SU(3)_c\times \SU(3)_N$) is required.

The Yukawa couplings of the RHNs to the SM lepton doublets, which are required to generate neutrino masses, explicitly break the $\SU(3)_N$ symmetry.
Therefore, unlike other constructions such as the Twin Higgs, $\SU(3)_N$ clearly cannot be an unbroken gauge symmetry at low energies. Instead, to be consistent with neutrino masses, $\SU(3)_N$ can be either a spontaneously broken gauge symmetry or an explicitly broken global symmetry. Since $\SU(3)_N$ is broken while $\SU(3)_c$ is unbroken (and gauged), quadratic divergences will appear at two loops, potentially threatening the naturalness of this scenario.  However, as we will argue below, these quadratic divergences lead to tunings which are tolerable, $\mathcal{O}(10\%)$, and can even be reduced by gauging the $\SU(3)_N$ symmetry and spontaneously breaking this symmetry at a scale close to the RHN mass. 

Regarding the neutrino sector, once we assume that $\SU(3)_N$ is broken, various Yukawa interactions and Majorana mass terms can be present in the Lagrangian. The Majorana mass terms softly break the global symmetry and do not introduce new quadratic divergences. The neutrino Yukawa couplings on the other hand represent a hard breaking of the global symmetry and will lead to quadratic divergences. However, provided these Yukawa couplings are not too large, these contributions will also be under control. This is analogous to the other light fermion Yukawa couplings, such as the bottom quarks, which do not jeopardize the naturalness of the theory. 
Motivated by these considerations, and the goal of generality, in this work we will allow all renormalizable terms, such as masses or Yukawa couplings, which explicitly break the required global symmetry but in such a way that the breaking does not spoil the naturalness of the theory any more than is already present due to the light fermion Yukawas. The approach taken here is thus a phenomenological bottom up exploration of the general class of models. 

The general features and bottom up requirements on the Natural Neutrinos scenario relating to naturalness are outlined in \Sec{sec:RHN}. 
In \Sec{sec:neut} we discuss the couplings necessary for the generation of neutrino masses, and in \Sec{sec:models} we describe several explicit low energy coset models which motivate the overall structure of the models.  Although we aim to focus on the phenomenology and low energy structure of this framework, we also comment on the embedding of these bottom up models into standard frameworks including composite and Twin Higgs scenarios.  In \Sec{sec:pheno} we will survey various aspects of the phenomenology of Natural Neutrinos beyond the generation of neutrino masses, including current constraints and future experimental prospects. We conclude in \Sec{sec:conclusions}. An Appendix provides a detailed discussion of proton decay and baryon number conservation in our scenarios.

\section{Right Handed Neutrinos as Top Partners}
\label{sec:RHN}

\begin{figure}[]
  \centering
  \includegraphics[width=0.452\textwidth]{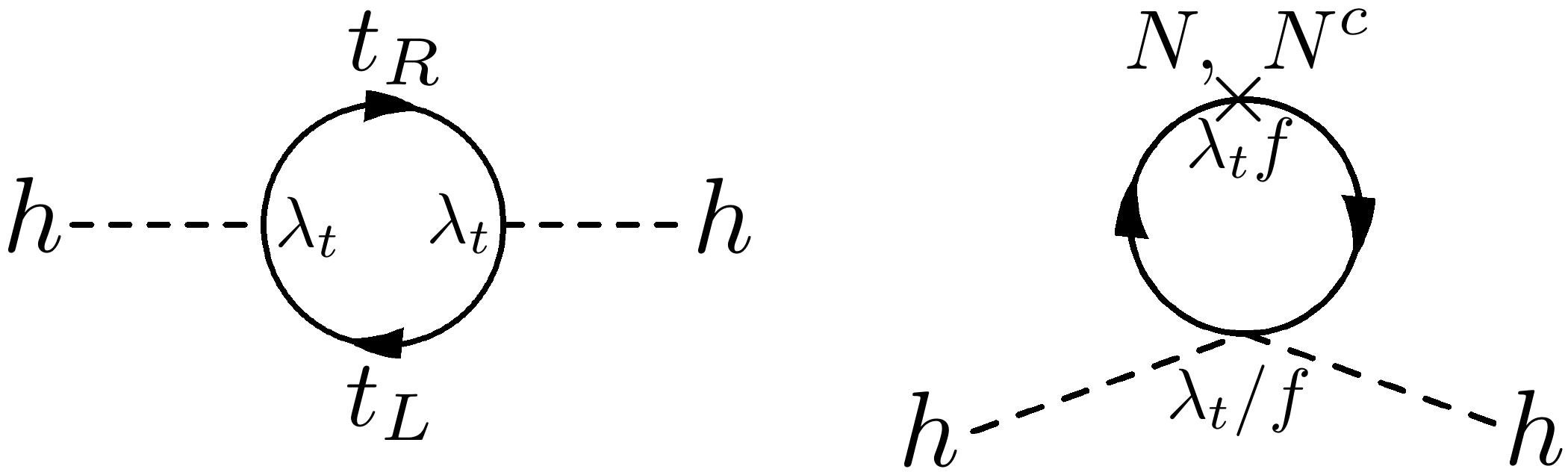}
  \caption{One-loop quadratically divergent corrections to the Higgs mass from top quarks and RHNs.  These contributions cancel in the proposed models as the coupling relation is enforced by a global symmetry.}
  \label{fig:cancel}
\end{figure}

We begin by describing the basic ingredients needed for the RHN to cancel the dominant top quark-induced quadratic divergence\footnote{In a full UV completion the Higgs mass is calculable, and the sensitivity to the cutoff $\Lambda$ is replaced by a sensitivity to the physical threshold, i.e., the masses of the states in the UV theory.}
of the Higgs mass, \ie, to be a top partner.  This is demonstrated schematically in \Fig{fig:cancel}.
Given that the RHNs are fermions, we are led to consider theories in which the Higgs arises as a pNGB, which provides a framework in which fermions can serve as top partners. In such theories, a large global symmetry $G$ is broken to a subgroup $H$ at a scale $f~\sim$ TeV, and the Higgs boson is identified as a pNGB of the coset $G/H$. The essential observation is that the RHN and the top quark can be joined in a multiplet transforming under $G$, such that, up to additional symmetry breaking effects, the top Yukawa coupling respects the global symmetry. In this way, the radiatively induced Higgs mass is governed by the mass of the RHN top partner, $m_N \approx \lambda_t f \lesssim$ TeV,  rather than the ultraviolet (UV) cutoff $\Lambda \sim  5-10$ TeV.

Let us expand on this idea. The low energy dynamics of the pNGBs are described by a non-linear sigma model, with the field
\be
\Sigma = e^{i \Pi/f} \Sigma_0, ~~~~~~ \Pi = \pi^a T^a,
\label{eq:sigma}
\ee
where $\pi^a$ are the pNGBs, $T^a$ are the broken generators of $G$, and $\langle \Sigma \rangle = |\Sigma_0| = f$. The minimal requirements on the symmetry breaking pattern are that the unbroken subgroup $H$ contains the electroweak group, $\SU(2)_W \times \U(1)_Y$, and that there are at least four pNGBs, ${\rm dim}\, G - {\rm dim}\, H
\geq 4$ in order to furnish a scalar electroweak doublet at low enegies. We will describe explicit coset constructions in Section~\ref{sec:models}. 

The third generation weak doublet quark $q$ and weak singlet top quark $t^c$ will be embedded in multiplets of $G$ which we denote $Q$ and $Q^c$, respectively. In addition to the top quarks, these multiplets contain neutral top partners $N$ and $N^c$, which we will identify with the RHNs. For this to occur, the $\SU(3)_c$ factor of the global symmetry must  be enlarged in order to accommodate neutral states. 
The simplest choices are $\SU(6) \supset \SU(3)_c \times \SU(3)_N$ or $\SU(3)_c \times \SU(3)_N$ with a $Z_2$ interchange symmetry, where we have identified $\SU(3)_N$ as the flavor symmetry of the right handed neutrinos. 
It is instructive to write a ``simplified model'' for the top quark Yukawa coupling, which can be realized in explicit  $G/H$ cosets (see Sec.~\ref{sec:models}):
\begin{eqnarray}
\label{eq:simplified}
 {\cal L} 
&=& \lambda_t Q \Sigma Q^c +\hc  \\
&=& \lambda_t \left[ q^A h t^c_A  + f \left(1-  \frac{h^\dag h}{2f^2}\right) N^i N^c_i + \dots \right]+\hc, \nn
\end{eqnarray}
where in the second line we have expanded out the $G$ multiplets $Q,Q^c$, and $\Sigma$ in their component fields to $\order(h^\dag h)$. In Eq.~(\ref{eq:simplified}), the $\SU(3)_c$ index $A = 1,2,3$ and the $\SU(3)_N$ index $i = 1,2,3$. 
As we will see in Sec.~\ref{sec:models}, in concrete realizations the precise structure of the interactions above can be generalized to include interactions with 
$\SU(2)_W\times \U(1)_Y$ charged states, and with different values of the coefficients in front of the couplings, but this simplified example will serve to illustrate the basic features relating to naturalness in this framework. 

With the additional coupling of the Higgs to the RHNs in Eq.~(\ref{eq:simplified}), the radiative contribution to the Higgs mass parameter, $\mu^2$, is softened from a quadratic to logarithmic sensitivity,
\begin{eqnarray}
\delta \mu^2 
& \simeq & -\frac{3 \lambda_t^2}{8 \pi^2 }\lambda_t^2 f^2 \log\frac{\Lambda^2}{\lambda_t^2 f^2}.
\label{eq:toploop}
\end{eqnarray}
This is also depicted in \Fig{fig:cancel}.  A naive estimate of the tuning in this theory is given by $|2\, \delta \mu^2/m_h^2|^{-1}$, which is of order $10\%$ for $f \sim 700$ GeV and $\Lambda \sim 5$ TeV. Without the RHN top partners, the radiative correction to the Higgs mass parameter is quadratically sensitive to the UV cutoff, leading to a tuning at the sub-percent level for the same choices of $f$ and $\Lambda$. In Eq.~(\ref{eq:toploop}) and below we evaluate the dimensionless couplings at the UV scale $\Lambda$ as is suggested by a renormalization group-improved analysis; see the discussion in Ref.~\cite{Craig:2015pha}.

To generate neutrino masses we must ultimately break the $\SU(3)_N$ symmetry. The implementation and breaking of the $\SU(3)_N$ symmetry is also relevant for tuning, as has been emphasized recently in Ref.~\cite{Craig:2015pha}.  If we take the $\SU(3)_N$ symmetry as a global symmetry then the quadratically divergent two-loop corrections to the Higgs mass coming from loops of top quarks which include $\SU(3)_c$ gluon exchange are not cancelled. This correction is given by 
\begin{eqnarray}
\delta \mu^2 & \simeq & - \frac{3 \lambda_t^2 g_3^2 }{8 \pi^4} \Lambda^2,
\label{eq:gluon-loop}
\end{eqnarray}
which, for a UV scale $\Lambda \sim 5$ TeV, yields a tuning of order $10\%$. 
 Beyond this correction, parametrically similar two-loop contributions arise from the fact that the Yukawa couplings to top quarks and top partners run differently due to the absence of $\SU(3)_N$ gluons.  In \cite{Craig:2015pha} it was estimated that the case of a global $\SU(3)_N$ symmetry would correspond to a tuning of $\mathcal{O}(10 \%)$ for $\Lambda \sim 5$ TeV.  Thus in our basic bottom-up approach for now we will simply assume that $\SU(3)_N$ is a global symmetry and allow this level of tuning.
 We will also allow explicit $\SU(3)_N$ symmetry breaking terms which lead to the generation of neutrino masses, restricting the sizes of such terms in such a way that does not increase the tuning further.  

We consider a tuning of $\mathcal{O}(10 \%)$ to be acceptable for the purposes of this work and in the following sections focus on the global symmetry limit. However, we note that it is possible to reduce this tuning further by instead considering the case of a spontaneously broken $\SU(3)_N$ gauge symmetry, with gauge coupling $g_3^N (\Lambda) \approx g^c_3 (\Lambda)$. 
In this case, the two-loop top-gluon corrections will be largely compensated by symmetry-related contributions coming from the top partners and $\SU(3)_N$ gluons, softening the UV sensitivity of the Higgs mass parameter to the scale of $\SU(3)_N$ breaking. If this scale is not too far from the $G\rightarrow H$ scale $f \sim$ TeV, the tuning will be appreciably reduced. This approach brings with it additional model-building questions related to the sector responsible for the breaking of the $\SU(3)_N$ gauge symmetry, which, while interesting, will not be pursued further in this work. 

There are a number of additional considerations required for a viable, natural theory of electroweak symmetry breaking. Besides the correction from the top loop, there is an additional important contribution to the Higgs mass parameter from the weak gauge boson loops, which can dominate over that of Eq.~(\ref{eq:toploop}) depending on the scales $f$ and $\Lambda$. This loop can be regulated with additional gauge boson partners, as occurs in Little Higgs or Twin Higgs theories.
However, whether or not such gauge partners are required depends again to some degree on the amount of tuning one is willing to accept, and in 
Section \ref{sec:models} we will consider example models both with and without gauge boson partners.
There is also the question of generating an appropriate scalar potential for the Higgs doublet, $h$. This occurs due to the explicit breaking of the global symmetry by the gauge and Yukawa interactions, but depends in detail on the low energy spectrum and to some extent the UV completion. As these issues are model dependent, we will set them aside for now and move next to the generation of neutrino masses in our scenario.

\section{Neutrino Masses}
\label{sec:neut}
 Through the top Yukawa interaction in Eq.~(\ref{eq:simplified})
the RHN top partner fields $N,N^c$ obtain $\SU(3)_N$-symmetric vector-like masses $M_N \sim y_t f$.  However once $\SU(3)_N$ and $\SU(3)_L$ (the flavor symmetry associated with the three $\SU(2)_W$ lepton doublets) are broken, which they must be in order to generate neutrino masses and mixings, 
a number of additional terms may arise consistent with the broken symmetries, including the spontaneously broken electroweak symmetry.  There are possible Majorana mass matrices from breaking of $\SU(3)_N$,
\be
\mathcal{L} \supset \frac{1}{2} \left( M^c_{M,ij} N^c_i N^c_j + M_{M,ij} N_i N_j \right)  ~,
\label{eq:Majorana}
\ee
and there are Dirac masses which require the breaking of $\SU(3)_N$, $\SU(3)_{L}$, and electroweak symmetry.  
These are
\be
\mathcal{L} \supset M^c_{D,ij} N^c_i \nu_j + M_{D,ij} N_i \nu_j   ~~.
\label{eq:Dirac}
\ee
Combining all terms, we may write the neutrino mass matrix in the $(\nu, N, N^c)$ basis as
\begin{equation}
{\cal M} = \left(    
\begin{array}{ccc}
0 & M_D & M_D^c \\
M_D^T & M_M &  M_N \\
M_D^{c \,T} &  M_N^T & M_M^c
\end{array}
\right),
\label{eq:mass-matrix}
\end{equation}
where each entry is a $3\times 3$ matrix.  
Since the vector-like RHN mass $M_N$ is $\SU(3)_N$ symmetric, while the other terms arise only after $\SU(3)_N$ symmetry breaking, we expect those terms to be suppressed in comparison to $M_N$. Furthermore, the mass $M_N \sim \lambda_t f$ is predicted to be below the TeV-scale by naturalness arguments. Therefore, the hypothesis of RHN top partners robustly predicts a TeV-scale seesaw mechanism!

We now explore the conditions required for the generation of the light neutrino masses. 
The determinant of the mass matrix (\ref{eq:mass-matrix}) is given by \cite{2011arXiv1112.4379P}
\begin{eqnarray}
\label{eq:big-det}
| {\cal M} |   & = &   | M_N | \times | M^T_N- M_M^c  M^{-1}_N M_M | \\ 
& & \times  |M_D^c  M^{-1}_N M_D + (M_D - M_D^c  M^{-1}_N M_M) \nonumber  \\ && \times  ( M^T_N \! - \! M_M^c  M^{-1}_N M_M)  ({M^c_D}^T \! - \! M_M^c  M^{-1}_N M_D^T)|, \nonumber 
\end{eqnarray}
where standard matrix multiplication is understood within the individual determinants.  This can be understood schematically from the determinant in the case of  a single generation,
\begin{eqnarray}
|{\cal M}|  & \sim & M_D (M_M^c M_D - M_N M_D^c) \\
& & +M^c_D (M^c_D M_M- M_N M_D) ~~. \nonumber 
\label{eq:singledet}
\end{eqnarray}
Alternatively, one can integrate out the heavy RHNs, which leads to the following mass matrix for the light neutrinos
\begin{equation}
{\cal M}_\nu = - 
\left( 
\begin{array}{cc}
M_D & M_D^c
\end{array}
\right)
\left( 
\begin{array}{cc}
M_M & M_N \\
M_N^T & M_M^c
\end{array}
\right)^{-1}
\left( 
\begin{array}{c}
M_D^T \\ M_D^{c\, T}
\end{array}
\right).
\end{equation}
It is clear that it is a collective symmetry breaking which leads to the generation of non-zero light neutrino masses.  In order to generate neutrino masses at least one of the Dirac mass terms of \Eq{eq:Dirac} must be non-zero and it is necessary that terms involving both $N^c$ and $N$ are present in order that lepton number is broken collectively.  The minimal options for generating neutrino masses are to have non-vanishing entries in the pairs of matrices shown in \Tab{tab:neutmass}, and we also display the approximate value of the light neutrino masses generated.  Alternative options such as non-vanishing $(M_N,M_N^c )$, $(M_D,M_N )$, or $(M_D^c,M_N^c )$ are not sufficient.  The former case is obvious as there is no coupling to the left-handed neutrinos.  The latter cases arise essentially due to a rank condition. It is interesting to note that the first option in Table~\ref{tab:collective} is similar to the
 so-called `Linear' \cite{Malinsky:2005bi} seesaw while the second and third option provide examples of the 
  `Inverse' \cite{Mohapatra:1986bd} seesaw (see also Refs.~\cite{Dev:2009aw,Dev:2012sg,Law:2012mj,Law:2013gma} for variant Inverse see saw scenarios). One particularly novel aspect of the collective breaking of lepton number is that it allows for comparatively large neutrino Yukawa couplings, resulting in various exotic phenomenological consequences. We will explore this feature in detail in Section~\ref{sec:pheno}.

\begin{table}[h]
\caption{\label{tab:neutmass}  Minimum pairs of mass matrices required for neutrino mass generation.  The Linear seesaw \cite{Malinsky:2005bi} may be realized for a pair of Dirac mass matrices and the Inverse seesaw \cite{Mohapatra:1986bd} for one Dirac and one Majorana mass matrix.}
  \begin{tabular}{ c | c}
    Non-zero mass-terms & Approximate neutrino masses \\ \hline
    $M_D,M_D^c $ & $\sim M_D M_D^c/M_N$ \\
    $M_D,M_M^c$ & $\sim M_D^2M_M^c/M_N^2$ \\
    $M_M,M_D^c$ & $\sim {M^c_D}^2 M_M/M_N^2$ \\
  \end{tabular}
  \label{tab:collective}
\end{table}

Although this work is not concerned with the details of flavor structures in the SM, this class of neutrino mass models may also be interesting from the perspective of masses and mixing angles.  Large hierarchies of masses are observed in the charged lepton and quark sector and also in the quark sector there are apparent hierarchical structures in the mixing angles.  However, in the models presented in this work the neutrino mass eigenvalues and mixing angles arise as the product of at least two seemingly unrelated matrices, suggesting that the mass and flavor structure of the neutrino sector would be very different from the other fermions of the standard model.

Given that the neutrino mass terms in Eqs.~(\ref{eq:Majorana}),(\ref{eq:Dirac}) break the $\SU(3)_N$ symmetry, it is reasonable to imagine that they originate from non-renormalizable operators. In this case, an immediate question is why we do not simply write down the Weinberg operator $(L\cdot h)^2/\Lambda$ \cite{Weinberg:1979sa}.  Assuming that the couplings in Eqs.~(\ref{eq:Majorana},\ref{eq:Dirac}) are the only spurions breaking $\SU(3)_N$ and $\U(1)_L$, then clearly any light neutrino mass must be proportional to the specific products of the spurions listed in Table~(\ref{tab:collective}). However, the scale suppressing these spurions may be much higher than $m_N$, corresponding to the UV dynamics that generate Eqs.~(\ref{eq:Majorana},\ref{eq:Dirac}), justifying our neglect of a bare Weinberg operator. 

In specific models, there may be further symmetries which constrain the form of the neutrino mass terms in Eqs.~(\ref{eq:Majorana}),(\ref{eq:Dirac}). Notably, in models with gauge boson partners $\acute{\rm a}$ la Twin Higgs, there are additional gauge symmetries under which the top partners are charged. Thus, to generate some of the terms in Eqs.~(\ref{eq:Majorana}),(\ref{eq:Dirac}) requires additional insertions of the fields which spontaneously break this symmetry. From the bottom-up approach we are pursuing here, there is no obstacle in writing down such terms.  

The Dirac masses $M_D$, $M_D^c$ originate from Yukawa couplings of the form $y_\nu L h N$, etc. and thus represent a hard breaking of the global symmetry protecting the Higgs mass. This is also true of the other SM fermions, provided we do not embed them in a $G$ multiplet with their own partner fields. However, this hard breaking does not upset the naturalness of the theory provided the Yukawa couplings are small enough. The radiative correction to the Higgs mass parameter from these Yukawa couplings is  given by
\be
 \delta \mu^2 \simeq -\frac{3 y_\nu^2}{8 \pi^2}\Lambda^2 . 
\ee
For a UV cutoff $\Lambda \sim 5$ TeV, the tuning estimate is greater than $\order(10\%)$ for neutrino Yukawa couplings smaller than $y_\nu \lesssim 0.25$. The Majorana masses $M_M, M_M^c$, softly break the global symmetry $G$, and thus do not introduce new quadratic divergences. However, the Higgs mass parameter is still sensitive to these parameters through the physical mass of the RHN top partners, and therefore naturalness requires $M_M, M_M^c\lesssim$ TeV. We will impose these constraints on the size of the explicit $\SU(3)_N$ breaking parameters generating neutrino masses throughout this work.

\section{Models}
\label{sec:models}

With the basic framework for a RHN top partner set out, we now turn to explicit phenomenological models where the low energy dynamics of the Higgs and the top partners are embedded within a specific symmetry-breaking structure. We will focus on three minimal scenarios for a pNGB Higgs: a) $\SU(3)/\SU(2)$, b) the custodially symmetric $\SO(5)/\SO(4)$, and c) $\SU(4)/\SU(3)$. Our main purpose in this section is to exhibit possible embeddings of the RHNs within the top quark multiplets, which will enforce an interaction structure analogous to Eq.~(\ref{eq:simplified}) leading to the cancellation of quadratic divergences. We will not wed ourselves to any particular UV physics, with the hope that these models may find both strongly coupled completions (\eg\ composite Higgs) or perturbative completions (\eg\ SUSY). 
Furthermore, a detailed investigation of radiative electroweak symmetry breaking is beyond the scope of this work.  We delay a discussion of the phenomenological constraints related to the EWSB sector, such as Higgs couplings measurements and electroweak precision tests, as well as observable signatures to Sec.~\ref{sec:pheno}.

\subsection{$\SU(3)/\SU(2)$}
\label{subsec:SU3SU2}

The minimal coset that furnishes a doublet pNGB under $\SU(2)_W$ is $\SU(3)/\SU(2)$ and has been investigated on a number of occasions~\cite{Schmaltz:2002wx,Kaplan:2003uc,Contino:2003ve,Perelstein:2003wd,Schmaltz:2004de,Berger:2012ec}. 
As a scenario for neutral top partners, this model was proposed in Ref.~\cite{Poland:2008ev}, and we will largely follow their discussion here. 
The global symmetry is taken to be $G = \SU(6) \times \SU(3)_W$.  The $\SU(6)$ factor contains $\SU(3)_c$, which is gauged, and the $\SU(3)_N$ flavor symmetry of the RHN. The $\SU(3)_W$ factor is broken to the $\SU(2)_W$ subgroup, resulting in 5 pNGBs, 4 of which form an $\SU(2)_W$ doublet to be identified with the Higgs. We also note there is a gauge singlet pNGB, although it will not be important for our discussion.

The global symmetry breaking is induced by a scalar field, $\Sigma$, transforming as a $\mathbf{3}$ under $\SU(3)_W$, which acquires a vacuum expectation value 
$\Sigma_0 = (0,0,f)$. The pNGBs can be parameterized by the non-linear sigma field as in Eq.~(\ref{eq:sigma}),
with
\begin{equation}
\Pi = \pi^a T^a = 
\left( 
\begin{array}{ccc}
0 & 0 & h_1\\
0 & 0 & h_2\\
h_1^\dag & h_2^\dag & 0
\end{array}
\right) + \dots,
\label{eq:pion}
\end{equation}
with $T^a$ the broken generators of $\SU(3)_W$ and we have suppressed the singlet pNGB.
We may write the sigma field explicitly as
\begin{equation}
\Sigma  =  
\left( 
\begin{array}{c}
i h_1 \, \displaystyle{ \frac{\sin( |h|/f)}{ |h|/f}}\\
i h_2 \, \displaystyle{ \frac{\sin( |h|/f)}{|h|/f}} \\
f\, \displaystyle{ \cos( |h|/f)}
\end{array}
\right),
\end{equation}
where $|h|\equiv \sqrt{h^\dag h}$. 

We now introduce the top quark and their partners, the RHNs. 
Following Ref.~\cite{Poland:2008ev} we add two fields $Q$, $Q^c$, which transform
under $\SU(6) \times \SU(3)_W$ as $Q \sim (\mathbf{6},\mathbf{\bar 3})$,  $Q^c \sim (\mathbf{\bar 6},\mathbf{1})$. We write these multiplets as
\begin{equation}
Q = 
\left( 
\begin{array}{cccccc}
q^A &  0 \\
0 &  N^i  \\
\end{array}
    \right),
    ~~~~~~~~
    Q^c = 
\left( 
\begin{array}{cccccc}
t^c_A  &  N^c_i  \\
\end{array}
    \right).
   \label{eq:QQc} 
\end{equation}
The top Yukawa coupling is written in an $\SU(6) \times \SU(3)_W$ symmetric manner as
\begin{eqnarray}
{\cal L} 
&=& \lambda_t Q \Sigma Q^c  \\
&=& \lambda_t \left[ i q^A h t^c_A  + f \left(1-  \frac{h^\dag h}{2f^2}\right) N^i N^c_i + \dots \right]+\hc , \nn
\label{eq:top-SU3}
\end{eqnarray}
which precisely reproduces the structure of our ``simplified model" in Eq.~(\ref{eq:simplified}).

Notice that $Q$ is an incomplete multiplet under the global symmetry group $G$. 
Naively this is worrisome since for generic incomplete multiplets the global symmetry is explicitly broken, and the cancellation of quadratic divergences does not hold. However, in this case one can understand the cancellation as a result of ``twisting''~\cite{Poland:2008ev}: 
starting from a full multiplet under $\SU(3)_c\times \SU(3)_W$, we have twisted one component, $N$, from $\SU(3)_c$ to $\SU(3)_N$. In this way, the couplings are structured such that the pNGB Higgs mass is protected against quadratic divergences. One possible realization of such incomplete multiplets is from a compactified 5th dimension, in which zero entries in the multiplets $Q,Q^c$ in Eq.~(\ref{eq:QQc}) correspond to would-be zero modes projected out by boundary conditions. 

Hypercharge can be accommodated in a straightforward manner by enlarging the global symmetry to $\U(6) \times \U(3)$, such that $Y = Y_6 +Y_3$, with $Y_6 = {\rm diag}(\tfrac{2}{3},\tfrac{2}{3},\tfrac{2}{3},0,0,0)$ is a $\U(6)$ generator not contained in the $\SU(3)_c$ subgroup and $Y_3 = {\rm diag}(\tfrac{1}{2},\tfrac{1}{2},0)$ is a $\U(3)$ generator not contained in the $\SU(2)_W$ subgroup. 
Finally, it is clear that this model, while being minimal, does not afford the possibility of having gauge boson partners and therefore the quadratic divergence coming from $\SU(2)_W$ gauge bosons is not cancelled.

\subsection{$\SO(5)/\SO(4)$}
\label{subsec:SO5SO4}
In this model the symmetry breaking pattern is $\SO(5)/\SO(4)$, which furnishes exactly four pNGBs which transform as a $(\mathbf{2},\mathbf{2})$ under the unbroken $\SU(2)_L\times \SU(2)_R \sim \SO(4)$ symmetry and are identified with the Higgs~\cite{Agashe:2004rs}. In the context of composite Higgs theories, this scenario is attractive due to the custodial symmetry of the strong sector, which protects against large contributions to the $\rho$ parameter. As before, we also assume an \SU(6) global symmetry containing $\SU(3)_c$ and $\SU(3)_N$. Note that in this scenario, as in the previous one, there is no possibility of neutral gauge boson partners, in contrast to the recent proposals of Refs.~\cite{Geller:2014kta,Barbieri:2015lqa,Low:2015nqa} based on the coset $\SO(8)/\SO(7)$. See also Ref.~\cite{Carmona:2014iwa} for pNGB composite Higgs models exploring the interplay between top partners, the lepton sector, and naturalness.

The global symmetry is broken by a scalar field, $\Sigma$, in the $\mathbf{5}$ representation of $SO(5)$, which acquires a vacuum expectation value $\Sigma_0 = (0,0,0,0,f)$. The Goldstones are parameterized as
\begin{equation}
\Sigma = \Sigma_0 e^{-i \Pi/f},   ~~~ \Pi = \sqrt{2} h^a T^{a},
\end{equation}
where $T^{a}$ are the broken generators. Explicit expressions for the generators are found in Ref.~\cite{Agashe:2004rs}. Using these generators, one obtains the expression
\begin{equation}
\Sigma =\frac{ \sin{|h|/f}}{|h|/f}
 \left(
\begin{array}{c}
h_1 \\
h_2 \\
h_3 \\
h_4 \\
|h| \cot |h|/f 
\end{array}
\right),
\end{equation}
where $|h| = \sqrt{h_1^2+h_2^2 + h_3^2 + h_4^2} = \sqrt{2 \, h^\dag h}$ 

We now consider the top sector. We embed the left handed top and bottom into $Q \sim (\mathbf{6}, \mathbf{\bar 5})$ and the right-handed top into $Q^c \sim (\mathbf{\bar 6},1)$. Each of these multiplets also contain top partners. The explicit embedding is 
\begin{equation}
\!\!\! Q  =  \frac{1}{\sqrt{2}} 
\left( 
\begin{array}{ccccc}
b & -i b & t & i t & 0 \\
{\cal E} & i {\cal E} & {\cal N} & - i {\cal N} & \sqrt{2} N
\end{array}
\right), ~~
Q^c= \left( 
\begin{array}{cc}
t^c  & N^c
\end{array}\right),
\end{equation}
where a complete $\SO(5)$ multiplet has been ``twisted'' between the upper and lower rows in $Q$, in analogy with the $\SU(3)/\SU(2)$ model described above. Therefore, while $Q$ is an incomplete multiplet under the full global symmetry, the top quark Yukawa and top partner-Higgs couplings are structured so that the pNGB Higgs mass is protected.

The top Yukawa is written as 
\begin{eqnarray}
{\cal L}  & \supset & \lambda_t Q \Sigma Q^c + \hc \\
 & \supset &  \lambda_t  \big[  q^A h t^c_A   +    \hat L^i h^\dag N^c_i +  f \left(1 - \frac{h^\dag h}{f^2} \right) N^i N^c_i  \big] + \hc \nn
 \label{eq:top-SO5}
\end{eqnarray}
In comparison to the model of Sec.~\ref{subsec:SU3SU2}
based on $\SU(3)/\SU(2)$, this model contains an additional electroweak doublet of fermions, $\hat L \sim (\mathbf{1}, \mathbf{2}, \tfrac{1}{2}) = ({\cal E}, {\cal N})$ (to be distinguished from the SM lepton doublet $L$). 

It is interesting to examine how the quadratic divergences cancel in this model. 
The second term in Eq.~(\ref{eq:top-SO5}) yields the same contribution to the quadratic divergence as the top quark loop. However, the coupling of the $h^\dag h N N^c$ term is twice as large as the one in 
Eqs.~(\ref{eq:simplified}) and (\ref{eq:top-SU3}), and so the $N N^c$ loop neatly cancels the quadratic divergences. 

As it stands, some of the fermions are massless, but we can add mass terms to the Lagrangian, such as, \eg, 
$M_L \hat L \hat L^c$, where $\hat L^c = ({\cal N}^c, {\cal E}^c)$. These mass terms softly break the global symmetry, but naturalness will not be spoiled provided that these masses are not significantly larger than $f$. Note also that there are additional possibilities for neutrino mass terms in this model in comparison to the ``simplified model'', although the discussion there can be straightforwardly generalized to include these new terms. 

In this scenario, hypercharge can be realized as follows. The global symmetry is enlarged to $\U(6) \times \SO(5)$, such that $Y = Y_6 +T^3_R$, with $Y_6 ={\rm diag}(\tfrac{2}{3},\tfrac{2}{3},\tfrac{2}{3},0,0,0)$ is a $\U(6)$ generator not contained in the $\SU(3)_c$ subgroup and  $T_R^3$ is the generator associated with the $\SU(2)_R$ subgroup of $\SO(5)$ (see Ref. \cite{Agashe:2004rs} for the explicit form of this generator).  As in the $\SU(3)/\SU(2)$ model, this model does not contain gauge boson partners. 

\subsection{$\SU(4)/\SU(3)$}
\label{sec:twin}

Finally we discuss a model based on the Twin Higgs~\cite{Chacko:2005pe}, which was the first model to realize neutral top partners.  The global symmetry is taken to be $\SU(6) \times \SU(4)$ (or alternatively $[\SU(3)\times \SU(2)]^2$ with a $Z_2$ interchange symmetry).  The $\SU(4)$ symmetry is assumed to be broken to $\SU(3)$ by the vacuum expectation value of a scalar field in the  fundamental representation of $\SU(4)$, $\Sigma_0 = (0,0,0,f)$, yielding 7 pNGBs. The $\SU(2) \times \SU(2)$ subgroup of $\SU(4)$ is gauged such that 3 of the pNGBs are eaten by the Twin $\SU(2)$ gauge bosons. The remaining 4 pNGBs form a doublet under $\SU(2)_W$ and are identified with the Higgs boson.  The pNGBs can be parameterized as in Eq.~(\ref{eq:sigma}), with
\begin{equation}
\Pi = \pi^a T^a = 
\left( 
\begin{array}{cccc}
0 & 0 & 0 & h_1\\
0 & 0 & 0 & h_2\\
0 & 0 & 0 & 0\\
h_1^\dag & h_2^\dag & 0 & 0
\end{array}
\right) + \dots,
\label{eq:pion}
\end{equation}
with $T^a$ the broken generators of $\SU(4)$.
We can obtain the expression for $\Sigma$ by expanding the exponential in Eq.~(\ref{eq:sigma})
\begin{equation}
\Sigma  =  
\left( 
\begin{array}{c}
i h_1 \, \displaystyle{ \frac{\sin( |h|/f)}{ |h|/f}}\\
i h_2 \, \displaystyle{ \frac{\sin( |h|/f)}{|h|/f}} \\
0 \\
f\, \displaystyle{ \cos( |h|/f)}
\end{array}
\right),
\end{equation}
with $|h| = \sqrt{h^\dag h}$

The top quark and its partners are embedded into representations of $\SU(6) \times  \SU(4)$ as $Q \sim (\mathbf{6},\mathbf{\bar 4})$,  $Q^c \sim (\mathbf{\bar 6},\mathbf{1})$. These fields are written explicitly as 
\begin{equation}
Q = 
\left( 
\begin{array}{cccccc}
q^A &  0 \\
0 &  \hat L^i  \\
\end{array}
    \right),
    ~~~~~~~~
    Q^c = 
\left( 
\begin{array}{cccccc}
t^c_A  &  N^c_i  \\
\end{array}
    \right),
\end{equation}
where $\hat L = (\hat E, \hat N)$ (which are Twin-sector fields to be distinguished from the SM lepton doublet $L$) and $N^c$ are the neutral top partners.
The top Yukawa coupling is written in a $\SU(6) \times \SU(4)$ invariant way as
\begin{eqnarray}
{\cal L} 
&=& \lambda_t Q \Sigma Q^c  + \hc \\
&=& \lambda_t \left[ i q^A h t^c_A  + f \left(1-  \frac{h^\dag h}{2f^2}\right) N^i N^c_i + \dots \right]+\hc , \nn
\label{eq:top-SU3}
\end{eqnarray}
which provides another example that reproduces the ``simplified model" in Eq.~(\ref{eq:simplified}), ensuring the cancellation of quadratic divergences in the top sector of the theory. 

In the usual implementation of the Twin Higgs, the second $\SU(3)$ Twin color symmetry (here identified with $\SU(3)_N$) is gauged, and left unbroken. However, as we have described in detail in Sec.~\ref{sec:RHN}, we take this to be a global symmetry. However, we do gauge the twin $\SU(2)_W$ symmetry, and therefore the model predicts gauge boson partners which cancel the quadratic divergences coming from the SM weak gauge bosons. 

Generalizations of the original Twin Higgs scenario in the context of composite Higgs models or their holographic duals have recently been constructed \cite{Geller:2014kta,Barbieri:2015lqa,Low:2015nqa}. These models are based on the symmetry breaking pattern $\SO(8)/\SO(7)$, and provide scenarios in which there is custodial symmetry and also protection of the Higgs mass not only from loops of the light degrees of freedom, but also from resonances near the compositeness scale. See also the earlier discussion of $\SO(8)/\SO(7)$ symmetry breaking pattern in Refs.~\cite{Chacko:2005pe,Chacko:2005vw,Chacko:2005un,Batra:2008jy}.  It would be interesting to consider RHN neutrino top partners in these frameworks.

It should be noted, especially in the Twin Higgs scenario, that there may be other SM-neutral fields which could also play the role of RHN.  However, here the working assumption is that only the Twin Top quarks take this role.  The presence of the additional Twin sector fields would also typically lead to significant modifications of the RHN phenomenology by introducing additional decay chains and production mechanisms.  We will comment on these possibilities where appropriate in the next section.

\section{Phenomenology}
\label{sec:pheno}
We now consider the main phenomenological consequences of the Natural Neutrinos scenario.  Unless specifically noted, throughout we will assume that only the SM and RHN top partner fields are present.

\subsection{Neutrino oscillation data}

We begin by detailing how the neutrino oscillation data can be described in our framework. Given that naturalness dictates the large scale $M_N \sim y_t f \lesssim$ TeV while the other entries in (\ref{eq:mass-matrix}) can only result from $\SU(3)_N$ breaking, it is natural to expect the latter to be suppressed in comparison to $M_N$, leading robustly to the prediction of a TeV scale seesaw.  Furthermore, one novel aspect of this framework is the fact that at least two additional entries are needed to generate neutrino masses (see Table~\ref{tab:collective}), indicating that there is a collective breaking of lepton number. This allows in principle for some of the Yukawa couplings to be relatively large, leading to the possibility of additional novel phenomena correlated with neutrino mass generation, as we will describe below. 

As already mentioned, there are two interesting limits in the mass matrix (\ref{eq:mass-matrix}) which realize the Inverse seesaw  (when $M_D, M^c_M$ or $M_D^c$, $M_M$ are the only additional non-vanishing entries) and the Linear seesaw (when $M_D, M_D^c$ are non-vanishing and the Majorana masses vanish). 
For concreteness, we will specialize to the case of the Inverse seesaw in what follows, with the neutrino mass matrix 
\begin{equation}
{\cal M} = \left(    
\begin{array}{ccc}
0 & M_D & 0  \\
M_D^T & 0 & M_N \\
0 &  M_N^T & M_M^c
\end{array}
\right).
\label{eq:mass-matrix-ISS}
\end{equation}
In the limit $M_D, M_M^c \ll M_N$, the spectrum consists of three light neutrinos and three pairs of heavy pseudo-Dirac fermions. This can be easily seen in the limit of one generation, which upon diagonalization of (\ref{eq:mass-matrix-ISS}) leads to the eigenvalues:
\begin{eqnarray}
\label{eq:mnu} m_{\nu} & \approx & \frac{M_D^2 M_M^c}{M_N^2}, \\ 
\label{eq:mN} m_{N,\pm} & \approx &\pm \left(M_N + \frac{M_D^2}{2M_N}\right) + \frac{M_M^c}{2}, 
\end{eqnarray}
valid in the limit $M_D, M_M^c \ll M_N$. We observe the collective breaking manifests through the dependence of $m_\nu$ on both $M_D$ and $M_M^c$.  The heavy states are split by an amount $|m_{N,+}| - |m_{N,-}| \approx M_M^c$.  For a natural value $m_N \sim y_t f  \sim 700$ GeV, the light neutrino mass scale of order $m_\nu \sim 0.1$ eV can be obtained in the two extremes of parameter space: 1) $y_\nu \sim 0.1$, $M_M^c \sim 10$ keV and 2) $y_\nu \sim 3 \times  10^{-6}$, $M_M^c \sim 100$ GeV.

The current global fit to the standard three flavor oscillation scenario  for the normal (inverted) ordering yields the following ranges for the squared mass differences and PMNS mixing angles~\cite{Gonzalez-Garcia:2014bfa,nufit}:
\begin{eqnarray}
& \Delta m_{21}^2 \, [10^{-5}\, {\rm eV}^2]  =  7.50^{+0.19}_{-0.17}  ~~ (7.50^{+0.19}_{-0.17}), & \nonumber \\
&\!\!\!\!\!\! \Delta m_{31}^2\, (\Delta m_{32}^2) \, [10^{-3}\, {\rm eV}^2]    =  2.547^{+0.047}_{-0.047} ~~( -2.449^{+0.048}_{-0.047}),  &\nonumber \\
& \sin^2 \theta_{12} = 0.304^{+0.013}_{-0.012} ~~ ( 0.304^{+0.013}_{-0.012}), & \nonumber \\
& \sin^2 \theta_{23} = 0.452^{+0.052}_{-0.028}   ~~ (0.579^{+0.025}_{-0.037}), & \nonumber \\
& \sin^2 \theta_{13} = 0.0218^{+0.0010}_{-0.0010} ~~ ( 0.0219^{+0.0011}_{-0.0010}), & 
\label{eq:nudata}
\end{eqnarray}
while the CP-violating phases are currently unconstrained.  
After integrating out the heavy RHNs we obtain the effective mass matrix for the three light neutrinos:
\begin{equation}
{\cal M}_\nu \simeq  M_D^T M_N^{-1} M_M^c M_N^{-1} M_D.
\end{equation}
A convenient way to automatically reproduce the low energy data is to employ the Casas-Ibarra parameterization~\cite{Casas:2001sr} of the Dirac mass matrix (see also Ref.~\cite{Arganda:2014dta}), which for the case of the Inverse seesaw is 
\begin{equation}
M_D = M_N \, (M_M^c)^{-1/2} \, R  \, (m_\nu)^{1/2} \, U_{\rm PMNS}^\dag,
\label{eq:CI-ISS}
\end{equation}
where $R$ is in general a complex, orthognal matrix satisfying $R^T R = 1$, $m_\nu = {\rm diag}(m_{\nu_1},m_{\nu_2},m_{\nu_3})$ is the physical light neutrino mass matrix, and $U_{\rm PMNS}$ is the PMNS matrix. In our numerical results here and below, we will fix $m_\nu$, $U_{\rm PMNS}$ as allowed by the global data (\ref{eq:nudata}), restricting to the normal ordering for simplicity and scanning over the lightest neutrino mass $m_{\nu_1}$. We furthermore fix $m_N = 700$ GeV while scanning over $R$ and $M_M^c$. In the left panel of \Fig{fig:neutrino-scan} we show the results of the scan in the $\overline{M_{M}^c}  - \overline{M_D}$ plane, where the bar denotes the average value of the non-zero entries in the associated mass matrices. 
All points in the plot reproduce the low energy neutrino data presented in Eq.~(\ref{eq:nudata}) (the gray points are excluded by additional constraints to be discussed below). The black line represents the correlation of Eq.~(\ref{eq:mnu}), 
$\overline{M_D} =  \overline{M_N} \,\sqrt{\overline{m_\nu}\big/ \, \overline{M^c_M} } $; notice that the points cluster around the line as expected.

\subsection{Non-unitarity of the PMNS matrix}

A novel aspect of the neutrino mass models proposed here, and the Inverse seesaw structure in particular, is the possibility of large neutrino Yukawa couplings. Such large couplings can manifest as a violation of unitarity of the PMNS matrix, which leads to a host of physical consequences~\cite{Antusch:2006vwa,Basso:2013jka,Antusch:2014woa}.

In general, the $9\times9$ neutrino mass matrix in Eq.~(\ref{eq:mass-matrix}) is diagonalized by a unitary matrix, $U$, such that
\begin{equation}
{\cal M}^{\rm diag} = U^{T} {\cal M}\, U.
\end{equation}
In the presence of the heavy RHNs, the $3\times3$ sub-matrix describing the mixing of the three light neutrinos, $\widetilde U$, is no longer unitary ($\widetilde U$ is to be distinguished from the standard {\it unitary} PMNS matrix $U_{\rm PMNS}$). 

Such deviations from unitarity can show up in a number of measurements. For example, the prediction for the Fermi constant, $G_F$, which is extracted from the precise measurement of the muon lifetime, is altered from the standard case of a unitary PMNS matrix. 
The experimentally determined quantity, 
$G_\mu = 1.1663787(6) \times 10^{-5}$ GeV$^{-2}$~\cite{Agashe:2014kda}, 
is related to $G_F$ in this case as
\begin{equation}
G_\mu^2 = G_F^2 (\widetilde U \widetilde U^\dag)_{\mu\mu}  (\widetilde U \widetilde U^\dag)_{ee}.
\end{equation}
In the standard case of mixing between only 3 light neutrinos, PMNS unitarity guarantees 
$(\widetilde U \widetilde U^\dag)_{ij} = \delta_{ij}$, while in our scenario this is no longer the case. 

\begin{figure*}
 \centerline{\includegraphics[width=0.45\textwidth]{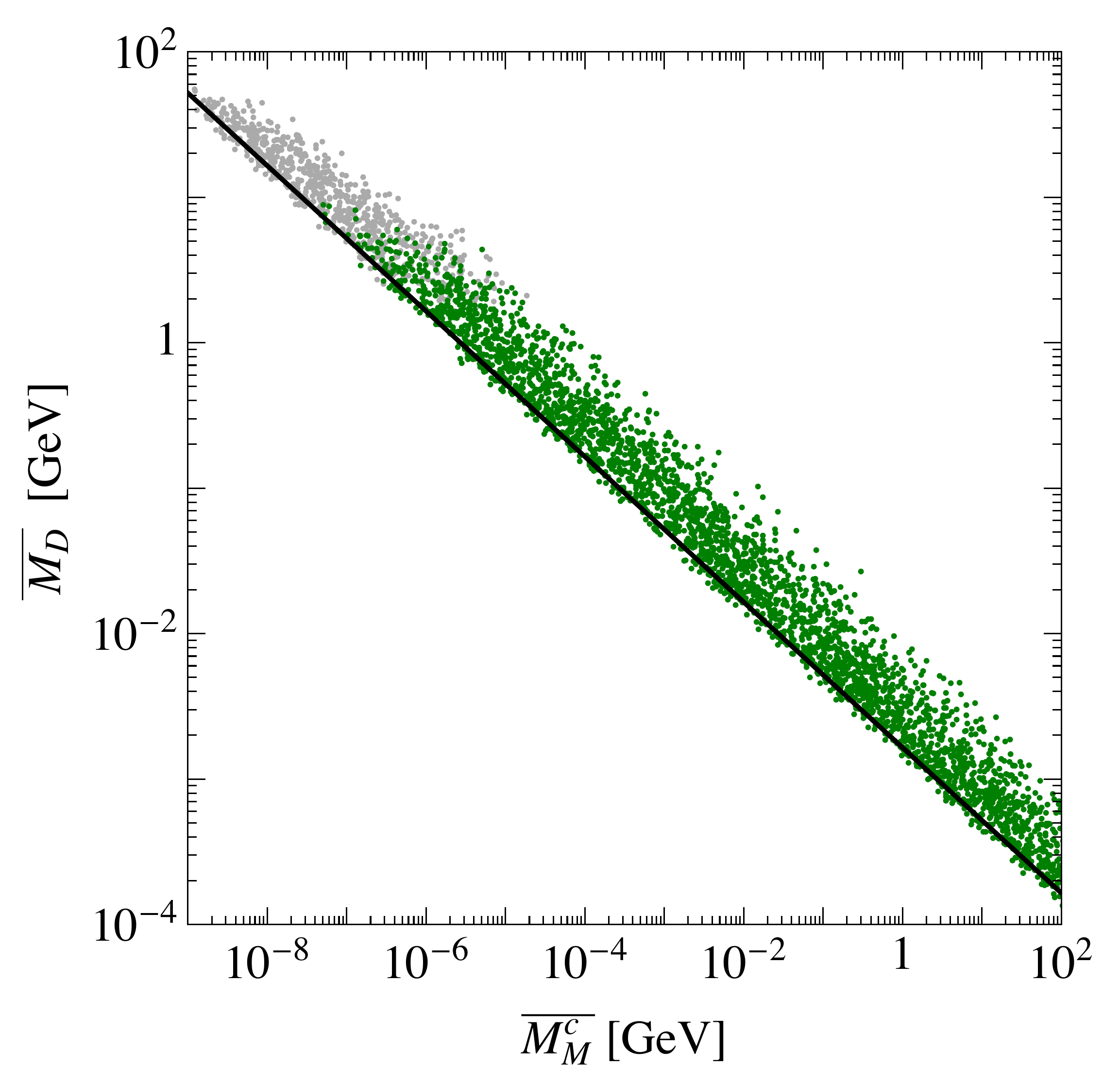} \hspace*{0.3cm} \includegraphics[width=0.45\textwidth]{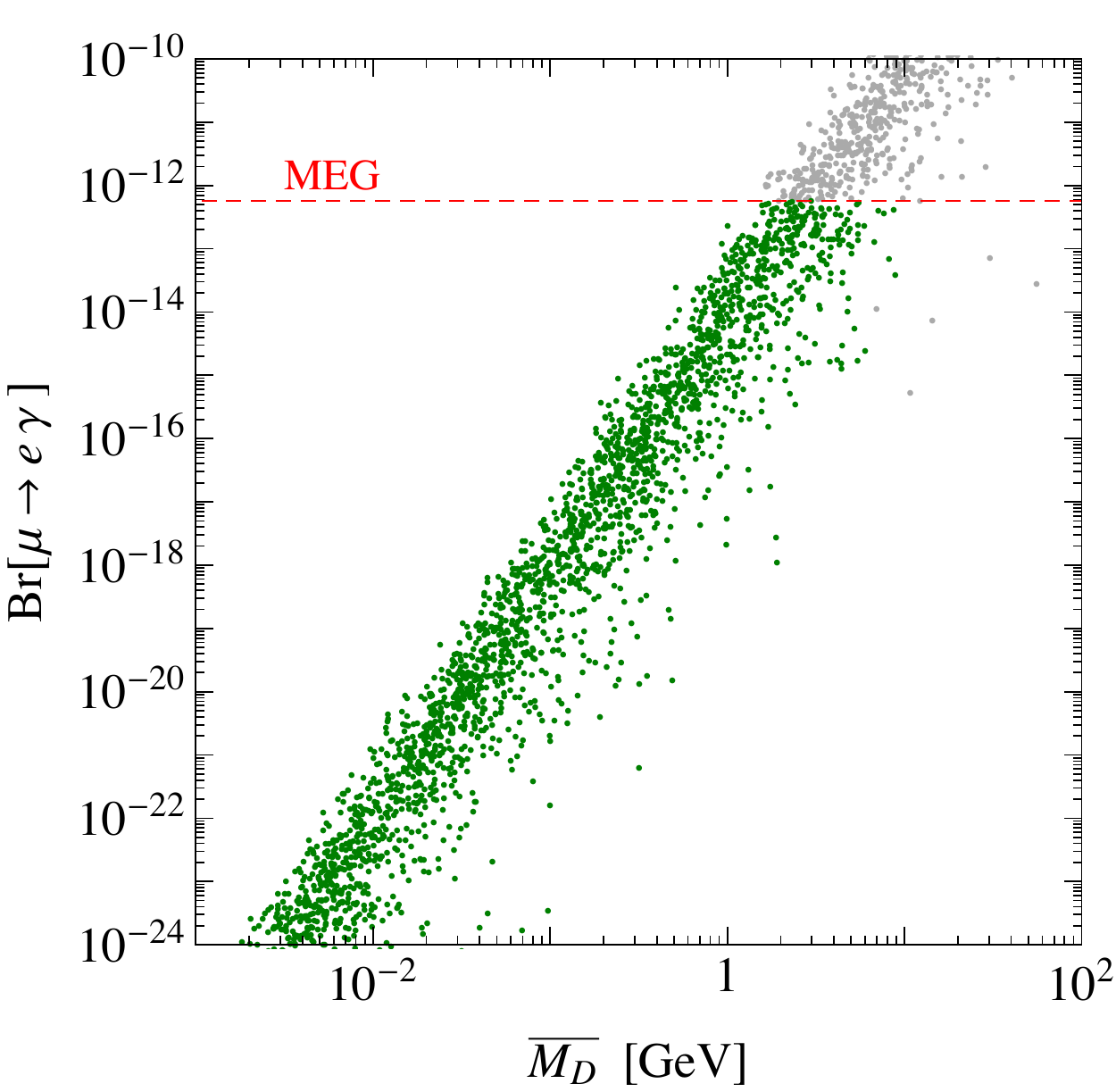}}
 \caption{
{\it Neutrino parameter space}: Here we display the results of a scan over the Majorana masses $M_M^c$ and the Dirac masses $M_D$ via the parameterization of 
Eq.~(\ref{eq:CI-ISS}). 
All points shown reproduce the low energy neutrino oscillation data in Eq.~(\ref{eq:nudata}), while the green points are also allowed by constraints on non-unitarity of the PMNS matrix~\cite{Antusch:2014woa} and lepton flavor violation.  
In the left panel we have displayed the results in the $\overline{M_{M}^c}-\overline{M_{D}}$ plane (the average Majorana and Dirac masses). The black line indicates the correlation expected from Eq.~(\ref{eq:mnu}).  The right plot shows the prediction for the lepton flavor violating decay $\mu \rightarrow e \gamma$ as a function of $\overline{M_{D}}$. We also display the current 90$\%$ C.L. limit from the MEG collaboration, ${\rm Br}(\mu\rightarrow e \gamma) < 5.7 \times 10^{-13}$~\cite{Adam:2013mnn}. 
}
 \label{fig:neutrino-scan}
\end{figure*}

Indeed, a suite of predictions for weak interaction observables are affected by the deviations from unitarity of $\widetilde U$, including $Z$ and $W$ boson decays, $Z$-pole asymmetries, invisible $Z$-boson width, the $W$-boson mass, weak mixing angle measurements, lepton flavor universality tests, lepton-flavor violating decays, and quark flavor CKM parameters. A recent study of these effects is presented in Ref.~\cite{Antusch:2014woa}, in which bounds are derived on the size of the deviations from unitarity on the quantities $(\widetilde U \widetilde U^\dag)_{ij}$, with $i,j = 1,2,3$. In particular, for the diagonal elements, we apply the following conservative $3\sigma$ C.L. limits~\cite{Antusch:2014woa}: 
\begin{eqnarray}
 1-(UU^\dag)_{ee} & < & 0.0018, \nonumber  \\ 
 1-(UU^\dag)_{\mu\mu} & < & 0.0007, \nonumber   \\  
   1-(UU^\dag)_{\tau\tau} & < & 0.005,    
\label{eq:non-unitarity-boundsa}
\end{eqnarray}
We note that these limits assume independent variations of the elements in the fits, while in our scenario, there will be correlations among the non-unitary parameters which are expected to change the precise limits. A full study of these effects are beyond the scope of this paper, but we expect the bounds in Eq.~(\ref{eq:non-unitarity-boundsa}) to be reasonably representative of the correlated ones in our scenario. 
Furthermore, there are limits on the off-diagonal (flavor-changing) elements, which we will return to below when we discuss lepton flavor violation. The green points in Figure~\ref{fig:neutrino-scan} are allowed by the constraints in Eq.~(\ref{eq:non-unitarity-boundsa}), while the gray points are excluded by a combination of these constraints and those coming from lepton flavor violation, as we discuss next.  Future possible lepton colliders such as the ILC, FCC-ee/TLEP, and CEPC, with their improvement in precision electroweak measurements, along with a suite of new low-energy experiments testing lepton universality, 
will probe deviations from PMNS unitarity at the 
$10^{-4} - 10^{-6}$ level~\cite{Antusch:2014woa}. 

Finally, it would be interesting to investigate further the effects of non-unitary mixing on neutrino oscillations, see for instance the recent stuy of Ref.~\cite{Escrihuela:2015wra}.

\subsection{Lepton flavor violation}

Another important consequence of the potentially large neutrino Yukawa couplings in our neutrino mass models are lepton flavor violating processes. A useful and comprehensive study of lepton flavor violating decays in seesaw models is given in Ref.~\cite{Ilakovac:1994kj} (see also Ref.~\cite{Deppisch:2004fa,Abada:2014kba}). Of particular importance is the $\mu \rightarrow e \gamma$ branching ratio, which places the strongest constraint on the neutrino Yukawa coupling in our scenario. The prediction for this branching ratio is~\cite{Ilakovac:1994kj}
\begin{equation}
{\rm Br}( \mu \rightarrow e \gamma) = \frac{\alpha_W^3 s_W^2}{256 \pi^2}\frac{m_\mu^5}{m_W^4 \Gamma_\mu}
\bigg\vert \sum_{i=1}^9  U_{\mu i} U^*_{e i}G\left(\frac{m^2_{N,i}}{m_W^2} \right)\bigg\vert^2,
\end{equation}
where the loop function is $G(x) = -(1/4)(1-x)^{-4}[(2x^3+5 x^2 - x)(1-x)+6x^3 \log{x}]$. 
Here, $m_{N,i}$ denote the 9 physical neutrino masses, and $m_\mu$ ($\Gamma_\mu$) is the muon mass (width), which can be found in Ref.~\cite{Agashe:2014kda}.
The strongest constraint on this branching ratio comes from the MEG collaboration,
\begin{equation}
{\rm Br}( \mu \rightarrow e \gamma)_{\rm MEG} < 5.7 \times 10^{-13}.
\end{equation}
In the right panel of Figure~\ref{fig:neutrino-scan} we display the prediction for this branching ratio as a function of the average Dirac mass. This measurement places the strongest constraint on the Dirac mass, requiring neutrino Yukawa couplings $y_\nu \lesssim  0.1$. 

Looking forward, there are exciting prospects for improving the sensitivity to LFV processes (see, \eg \cite{Albrecht:2013wet,bergertalk} for recent overviews). For $\mu\rightarrow e \gamma$, the MEG-II upgrade will be able to improve the branching ratio limit by roughly an 
order of magnitude~\cite{Gerone:2014nna}. Other prospects for LFV involving muons include experiments such as COMET, DeeMee, Mu2e, and PRISM, which are expected to improve the bound on the coherent muon-to-electron conversion rate by several orders of magnitude in the coming years~\cite{Albrecht:2013wet,bergertalk}. Likewise, the Mu3e experiment has the capability to improve the bound on the $\mu \rightarrow 3e$ branching ratio by four orders of magnitude, probing branching ratios down to the level of $10^{-16}$~\cite{Albrecht:2013wet,bergertalk}. There are also exciting prospects for $\tau$-LFV processes, which can be probed by LHCb and flavour factories such as Belle-II and Super KEKB~\cite{Albrecht:2013wet}, as well as LFV $Z$ decays which can be tested at future lepton colliders~\cite{Gerone:2014nna}.

\subsection{Neutrinoless double beta decay}
A classic signature of the Majorana nature of the neutrino is neutrino-less double beta decay. The amplitude for this process contains the sum (see \eg~\cite{Benes:2005hn,Mitra:2011qr,Pascoli:2013fiz,Faessler:2014kka}),
\begin{equation}
{\cal A}_{0\nu2\beta} \propto \sum_{i = 1}^9 U_{e i}^2 \frac{m_{N,i}}{\langle p^2 \rangle - m_{N,i}^2},
\label{eq:beta}
\end{equation}
where $\langle p^2 \rangle \sim (100 \, {\rm MeV})^2$ is the characteristic momentum scale of the process.
In principle the heavy neutrinos can enhance the rate of this process. In particular, neglecting the mixing angles for the moment, we observe from Eq.~(\ref{eq:beta}) that the contribution of the heavy neutrinos is enhanced in comparison to that of the light neutrinos by the factor ${\cal A}_N/{\cal A_\nu} \sim \langle p^2 \rangle/(m_\nu M_N) \sim 10^5$. Indeed, strong bounds have been placed on the size of {\it generic} mixing elements,  $|U_{ei}|^2 \lesssim 10^{-5}$,  for TeV-scale RHNs~\cite{Faessler:2014kka}. 

However, since the process violates lepton number, it is clear that the rate must be governed by the small collective breaking of lepton number. This is true even if the mixing angles are large, which can happen if $y_\nu$ is large. 
Working in the flavor basis, it is easy to see that the contribution of the heavy neutrinos requires two Yukawa insertions and one Majorana mass insertion, and therefore must be proportional to 
${\cal A}_N \sim M_D^2 M_M^c/M_N^4 \sim m_\nu/M_N^2$. 
This can also be seen in the mass basis, in which case there is a fine cancellation between the amplitudes of the split pseudo-Dirac states. Therefore, we conclude that the contribution from the heavy neutrinos is negligible. It is of course still possible that neutrinoless double beta decay is observable in this scenario, although it will be a consequence of the light neutrinos and in this sense is not different from the minimal scenario in which Majorana neutrino masses are described by the dimension 5 Weinberg operator. 

\subsection{Higgs couplings}
A generic prediction of pNGB Higgs scenarios is the modification of the Higgs couplings to SM gauge bosons and fermions, which are conventionally parameterized as \cite{Contino:2010mh} 
\begin{equation}
a = \frac{g_{hVV}}{g_{hVV}^{\rm SM}}, ~~~~ c = \frac{g_{hff}}{g_{hff}^{\rm SM}}, ~~~~
\label{eq:ac}
\end{equation}
which are a function the combination $v^2/f^2$. The precise form of $a,c$ depends on the symmetry breaking pattern and the embedding of the SM fermions in to $G$ representations, but are generically of the form 
$a,c \approx 1 - \order(v^2/f^2)$. 
Currently, the LHC $7+8$ TeV data are consistent with the SM values $a = c = 1$ and constrain the parameter $a$ $(c)$ at the $\pm10\,(40)\,\%$ level at 3$\sigma$ C.L.~\cite{ATLAS-CONF-2014-009,CMS-PAS-HIG-14-009}. Given that the precision on $a$ dominates, we can infer a constraint on the symmetry breaking scale $f \gtrsim 400$ GeV. Looking towards the future, the LHC will eventually be able to measure $a$ at the few percent level, probing $f \sim 1$ TeV.  Beyond this, future $e^+ e^-$ machines will have the capability to measure $a$ at the per-mille level, probing $f$ in the multi-TeV range~\cite{Dawson:2013bba}. 

As the RHN are neutral under the entire SM gauge group no additional corrections to the Higgs-gluon-gluon or Higgs-photon-photon couplings are expected. Finally we note that, in contrast to more general Inverse see saw models~\cite{BhupalDev:2012zg}, since the RHN top partners are constrained to be heavier than ${\cal O}(500\,{\rm GeV})$ there are no new Higgs decays directly into the new neutrino states.

\subsection{Electroweak precision tests}
A major constraint on the global symmetry breaking scale $f$ comes from the precision electroweak data.
There are several potential contributions to the oblique parameters, $S$ and $T$~\cite{Peskin:1991sw}. First, there is an irreducible contribution which arises due to the modification of the $hWW$ and $hZZ$ couplings, which in the models above can be described in terms of the parameter $a \sim 1 + {\cal O}(v^2/f^2)$ defined in Eq.~(\ref{eq:ac}).
This modification implies that, in contrast to the SM, the IR log divergences in the gauge boson vacuum polarizations do not completely cancel up to the scale $\Lambda$, leading to a contribution to $S$ and $T$~\cite{Barbieri:2007bh,Contino:2010rs}:
\begin{eqnarray}
\Delta S = \frac{1}{12 \pi}(1-a^2)\log\left( \frac{\Lambda^2}{m_h^2}\right), \nn \\
\Delta T = - \frac{3}{16 \pi c_W^2}(1-a^2)\log\left( \frac{\Lambda^2}{m_h^2}\right).
\label{eq:ST}
\end{eqnarray}
The corrections depend logarithmically on the UV cutoff. For instance, in the case of a strongly coupled UV completion, one expects $\Lambda \sim 4 \pi f = 4 \pi v/\sqrt{1-a^2}$. Using, for example, the most recent results of the Gfitter group, $S = 0.06 \pm 0.09$, $T = 0.1\pm 0.07$, with a correlation coefficient $\rho = 0.91$, we find that this places a strong bound $f \gtrsim 1$ TeV at 3$\sigma$ C.L.. However, it should be noted that the SM point ($S = T = 0$) lies outside the $1\sigma$ ellipse. In this light, a more conservative bound can be placed by measuring the deviation of $\Delta S$, $\Delta T$ in Eq.~(\ref{eq:ST}) from the central value rather than the SM point, in which case the bound of $f$ weakens to $f \gtrsim 500$ GeV. If instead one considers a perturbative completion the UV cutoff can be appreciably lower, reducing the shifts to the oblique parameters in Eq.~(\ref{eq:ST}).

Beyond these irreducible pieces, there can be other contributions to the oblique parameters depending on the UV completion. In strongly coupled UV completions, such as composite Higgs theories, one expects tree-level contributions from the resonances at the scale $m_\rho  \sim  g_\rho f$ where $g_\rho$ parameterizes the strength of the coupling of the resonances~\cite{Giudice:2007fh}. If the unbroken subgroup $H$ does not enjoy a custodial symmetry, one expects large contributions to the $T$ parameter of order $(1/\alpha)(v^2/f^2)$, constraining $f$ to be in the unnatural multi-TeV range. Provided this contribution is controlled by a custodial symmetry, one still expects a contribution to the $S$ parameter of order $\Delta S \sim  (4 \pi/g_\rho^2)(v^2/f^2)$, which depending on $g_\rho$ can be larger than the contribution in Eq.~(\ref{eq:ST}). However, in a weakly coupled UV completion, such as SUSY, these contributions are expected to be absent. 

Finally, there can be model-dependent contributions from other light states in the spectrum, such as the partners to the SM fields. Importantly, one can obtain a positive contribution to the $T$ parameter if the new states break custodial symmetry through their couplings to the Higgs, allowing to significantly relax the constraints on $f$. 

Possible future $e^+ e^-$ machines, such as the ILC, FCC-ee, and CEPC, will be able to determine the oblique parameters at the percent level~\cite{Fan:2014vta}, probing scales $f \sim 1$ TeV.

\subsection{Direct production at high energy colliders}
It is possible that the RHN top partners may be produced at colliders.  Before considering the possible production mechanisms it is worthwhile to first determine the decay channels. Here we will consider the minimal scenario is which the RHN top partners are the only new states in the spectrum. If the Yukawa couplings, $h N L$ and $h N^c L$, were set to zero the left-handed neutrino masses would vanish and the RHNs would become stable due to a $Z_2$ symmetry.  Thus in a complete model all decays must proceed through the Yukawa couplings.  Making use of Goldstone equivalence the possible decay chains of the heavy RHN are
\begin{eqnarray}
N, N^c & \to & h \nu \, , \nn \\ 
N, N^c & \to & Z \nu \, , \nn \\ 
N, N^c & \to & W^+ l, W^- \overline{l} \, . 
\end{eqnarray}
Thus, there are a variety of final states which may be useful for collider searches, in particular involving missing transverse energy (MET) and/or leptons resulting from the fact that decays proceed through lepton Yukawas.  Furthermore, decays of the heavy RHN will have a final state flavor structure which is correlated with the flavor structure of the light neutrino masses and mixings. Thus, in the charged current decays $N\to W l$ final states which are not democratic in lepton flavor may arise, giving clues as to the origin of neutrino masses and mixings at collider experiments. 

We emphasize again that here we are considering a minimal spectrum with the only new states being the RHN top partners. However, it should be kept in mind that if there are additional light neutral states then the RHN are likely to decay dominantly into those states.  This is particularly relevant if the model is embedded in a Twin Higgs scenario, such as the one described in Section~\ref{sec:twin}.  In particular, in this model the RHN (the ``Twin top'') may dominantly decay into a Twin $W$ boson and its Twin SU(2) fermion partner $E$ (the ``Twin bottom'').\footnote{We would like to thank Gian Giudice for emphasizing this point.}  There is a great deal of Twin sector spectrum dependence on the possible decay chains. However, we note that the Dirac masses of \Eq{eq:Dirac} may be the only couplings that mix Twin sector fermions with SM fermions, and thus even if there are long decay chains within the Twin sector they may terminate back into the visible sector through these operators, potentially leading to large multiplicity final states involving fermions. 
Furthermore, an additional experimental opportunity is present in these scenarios as the Higgs boson can decay to these light Twin sector states. In general, the absence of a confining $\SU(3)_N$ force suggests a distinct exotic Higgs decay phenomenology in comparison to usual Twin Higgs scenarios, such as that described recently in Ref.~\cite{Craig:2015pha}.

\subsubsection{Hadron colliders}
\begin{figure}[]
 \centering
 \includegraphics[width=0.4\textwidth]{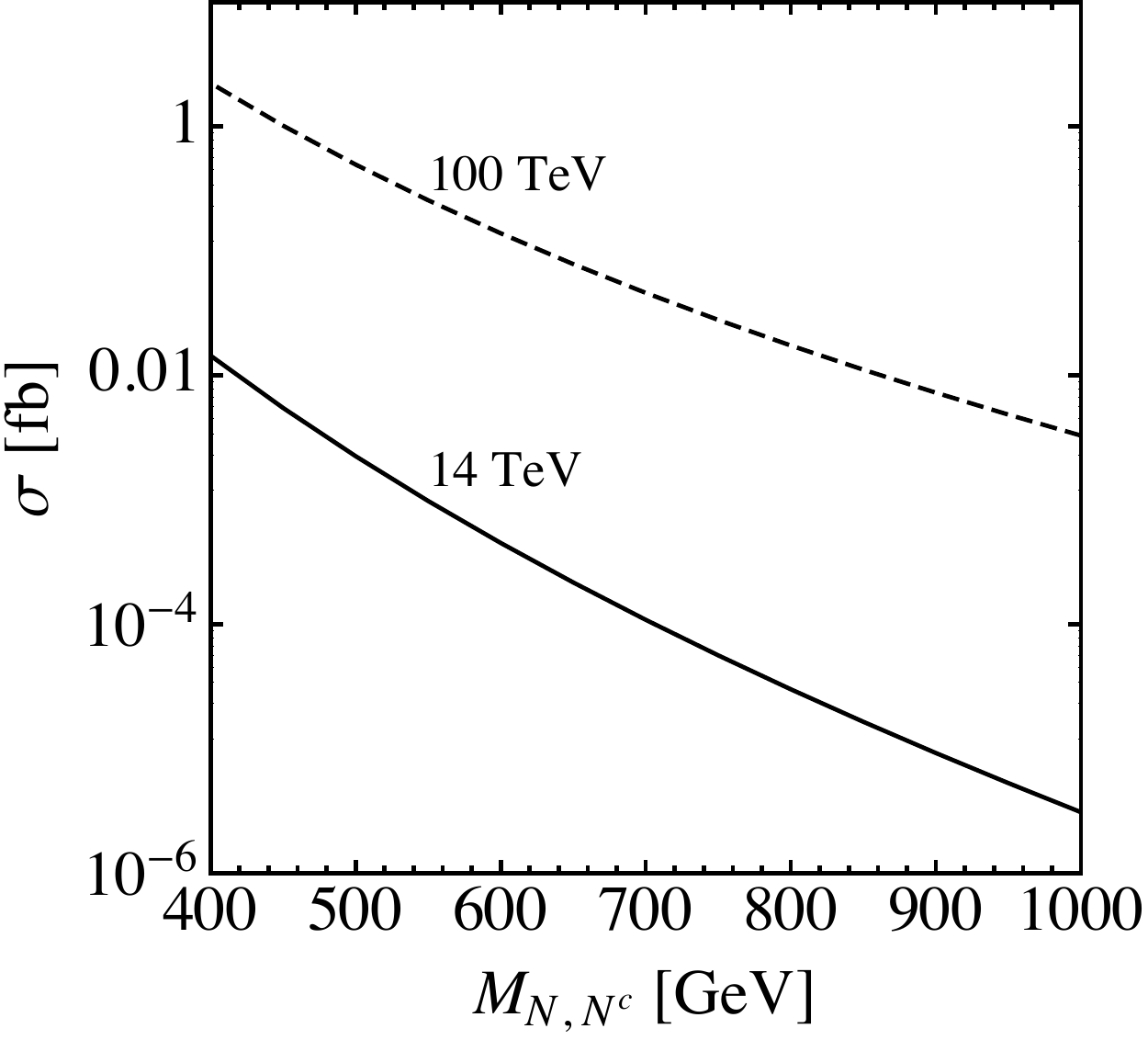}
 \caption{The cross section for pair producing the RHN at 14 and 100 TeV.  A vanishing Yukawa coupling has been chosen such that pair production only occurs through the Higgs coupling of \Eq{eq:simplified}.}\label{fig:pairprod}
\end{figure}

It is possible to produce the RHN top partners at hadron colliders.  In the limit of vanishing Yukawa couplings they may only be pair-produced, and this proceeds via gluon fusion $gg\to h^* \to N N^c$ due to the coupling of \Eq{eq:simplified}.  This production process is relatively model-independent as the coupling is dictated by the cancellation of quadratic divergences.  Assuming only the RHN and the SM the possible final states from RHN pair production are
\begin{eqnarray}
N N^c & \to & h h \nu \nu, \nonumber \\
N N^c & \to & h Z \nu \nu, \nonumber \\
N N^c & \to & Z Z \nu \nu, \nonumber \\
N N^c & \to & W l h \nu, \nonumber \\
N N^c & \to & W l Z \nu, \nonumber \\
N N^c & \to & W l W l \nu.
 \end{eqnarray}
In \Fig{fig:pairprod} we show the inclusive cross-section for pair production at the 14 TeV LHC, which was calculated by implementing the model in {\sc{FeynArts}}, and then using the {\sc{FeynArts}}, {\sc{FormCalc}}, and {\sc{LoopTools}} suite of packages \cite{Hahn:2000kx,Hahn:1998yk} to perform the tree-level and loop calculations.  These cross sections are typically small, hence at most $\mathcal{O} (\text{few})$ RHN pair production events would be expected with $300 \text{ fb}^{-1}$ of integrated luminosity at the LHC for small RHN mass.  The pair production cross section is significantly larger at 100 TeV suggesting a 100 TeV collider would be more promising for discovering RHN top partners.

\begin{figure}[]
 \centering
 \includegraphics[width=0.4\textwidth]{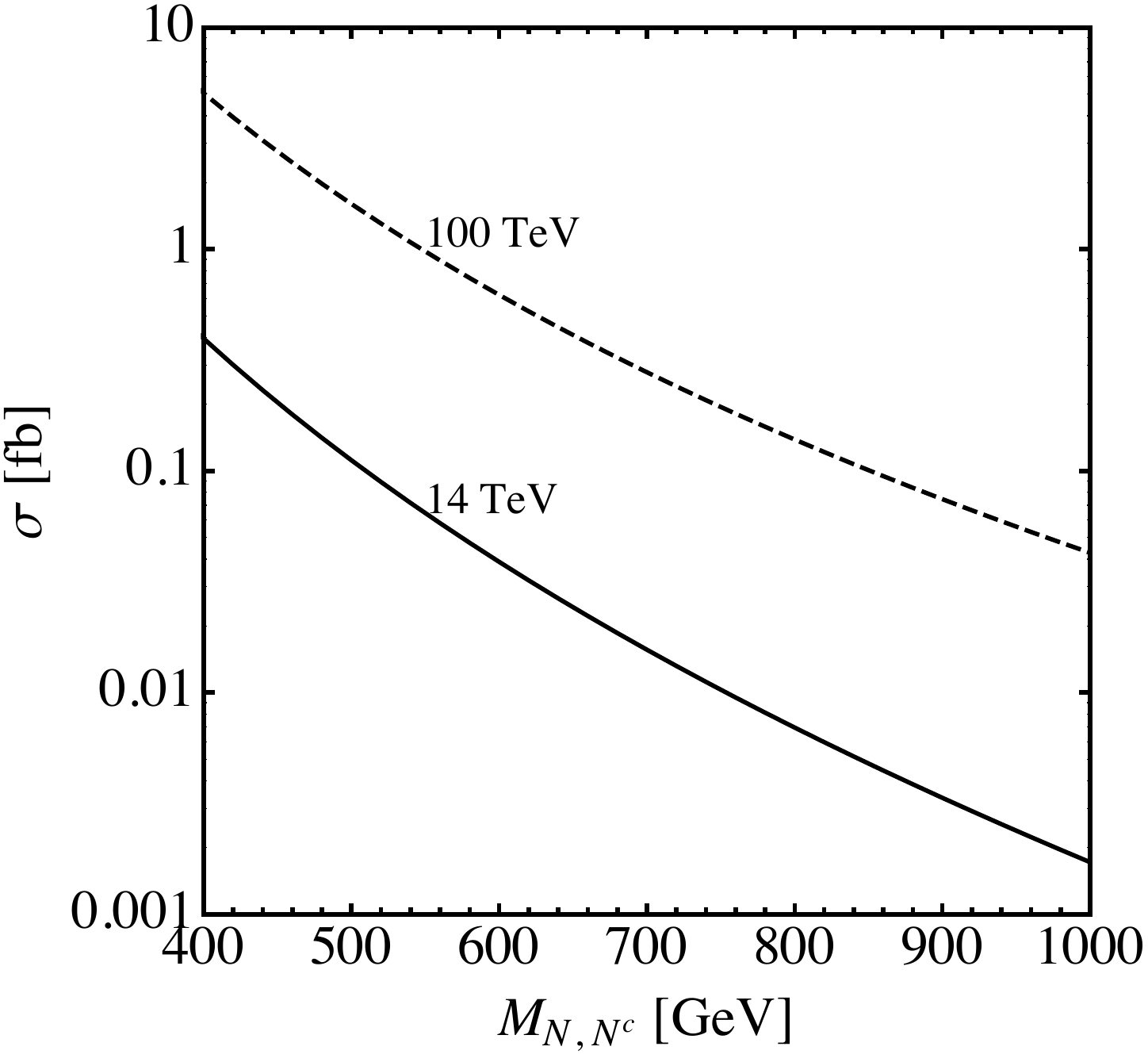}
 \caption{The cross section for producing a single RHN in association with a charge lepton, $p p \to N \overline{l}$, $p p \to N^c l$, at 14 and 100 TeV.  A Yukawa coupling of $Y = 0.1$ has been chosen.}\label{fig:singprod}
\end{figure}

In the electroweak symmetry breaking vacuum the Yukawa couplings mix the RHN with the left-handed neutrinos, thus it is possible to singly produce the RHNs in association with a $\nu$ (neutral current) or a charged lepton $l$ (charged current).  In \Fig{fig:singprod} we show typical cross sections for single RHN production in association with a charged lepton, for a relatively large Yukawa coupling $y_\nu=0.1$.\footnote{We have compared our calculations against  \cite{Das:2012ze} and find they are in good agreement.}  For typical RHN masses of $\mathcal{O}(700 \text{ GeV})$ only a handful of events would be expected with $300 \text{ fb}^{-1}$ of integrated luminosity.  The cross section increases by an order of magnitude when going from 14 to 100 TeV, again suggesting a 100 TeV proton-proton would be more promising for discovering the RHN top partners.  By considering the possible decay chains of the RHN the typical backgrounds to single production in association with a neutral or charged lepton would be the di-boson production processes
\begin{eqnarray}
& \text{SIG: } Z^* \to \nu  N  \to  h \nu \nu  \qquad & \text{BG: } h  Z(Z \to \nu \nu), \nonumber \\
& \text{SIG: } Z^* \to \nu  N  \to  Z \nu \nu  \qquad & \text{BG: } Z  Z(Z \to \nu \nu), \nonumber  \\
& \text{SIG: } Z^* \to \nu  N  \to  W l \nu  \qquad & \text{BG: } W  W(W \to l \nu), \nonumber  \\
& \text{SIG: } W^* \to l  N  \to  h l \nu  \qquad & \text{BG: } h  W(W \to l \nu), \nonumber  \\
& \text{SIG: } W^* \to l  N  \to  Z l \nu  \qquad & \text{BG: } Z  W(W \to l \nu), \nonumber  \\
& \text{SIG: } W^* \to l  N  \to  W l l  \qquad & \text{BG: } W  Z(Z \to l \overline{l}). 
 \end{eqnarray}
By taking into account the typical cross sections for the background processes it would appear that detecting the RHN top partners may be challenging at the LHC, although it is difficult to assess the possibilities adequately without performing a full collider study with optimized cuts.  A related study in the context of the Inverse seesaw embedded within the NMSSM finds that it may be possible to detect the heavy RHN at the LHC in the trilepton+MET final state with suitable cuts applied \cite{Das:2012ze}, however this was for $M_N = 100$ GeV which has a much larger production cross section. We note that striking collider signatures of lepton-number violation~\cite{Dicus:1991fk,Datta:1993nm,Han:2006ip,Atre:2009rg,Dev:2013wba,Alva:2014gxa,Das:2014jxa,Deppisch:2015qwa} are suppressed in our scenario by the small collective breaking of lepton number. 

Finally, the presence of additional heavy colored states in the multi-TeV range may be expected in a UV completion. Although access to these states at the LHC will be challenging it may be possible at a future 100 TeV hadron collider, and this possibility deserves further study.

\subsubsection{$e^+ e^-$ colliders}
Besides the powerful indirect constraints which could be achieved with future $e^+ e^-$ colliders it is worth considering the direct constraints which may be possible.  Due to the expectation that $v/f \ll 1$ the RHN masses are expected to be $M_N \gtrsim 400$ GeV, thus pair production of RHN would require at least a $1$ TeV $e^+ e^-$ collider such as the ILC, and if the masses were $M_N \gtrsim 700$ GeV then a significant increase in CM energy would be required for pair production.  On the other hand RHNs could be singly produced through $e^+ e^- \to Z^* \to \nu N$ and a $1$ TeV $e^+ e^-$ collider such as the $1$ TeV option of the ILC may be able to probe a significant portion of the relevant parameter space.  A study for lighter RHN at the ILC was performed in \cite{Das:2012ze} (see also \cite{Banerjee:2015gca}) and it was found that for $M_N = 150$ GeV a statistically significant excess may be observable.  It would be interesting to perform a similar study for the heavier RHN to determine the 
ty of \eg\ the ILC to Natural Neutrinos.

\section{Outlook}
\label{sec:conclusions}

We now speculate on some aspects that have not been considered or developed in this work, but which may lead to promising model-building and phenomenological avenues of future development.

\subsection{Neutrino Mass Models}
Our approach to the $\SU(3)_N$ and $\U(1)_L$ breaking terms that generate neutrino masses in this work has been bottom-up, allowing all terms at the renormalizable level consistent with naturalness constraints. It would be very interesting to explore explicit models that generate these couplings. Depending on how the $\SU(3)_N$ symmetry is implemented, \ie, as a global or spontaneously broken gauge symmetry, this will involve fields responsible for the breaking of $\SU(3)_N$ as well as new states mediating the couplings between the SM lepton doublets, the pNGB Higgs, and the RHN top partners. Explicit constructions of this type are likely to give guidance to the expected sizes of the effective neutrino Yukawa couplings and Majorana masses, and may lead to novel phenomenology.  It is worth noting that the terms needed for neutrino mass generation are small, thus they may be generated by new fields well above the UV scale $\Lambda$, or at $\Lambda$ but with suppressed couplings.

\subsection{UV Completion}
The couplings and field content in this work were motivated by symmetry considerations which ensured the cancellation of quadratic divergences due to the IR degrees of freedom alone. Ultimately it will be desirable to construct a full UV theory with a calculable Higgs mass that dynamically accounts for the origin of $\SU(3)_N$ and $\U(1)_L$ breaking.   
In such a UV completion, there are likely to be additional naturalness considerations that go beyond the na\"{\i}ve bottom-up coupling structures of \Sec{sec:models}.
Thus, although the low energy structure of Natural Neutrino models appears relatively uncomplicated, this does not necessarily imply that realizing a fully UV complete model would be straightforward and work towards this goal is necessary to put Natural Neutrino models on a firmer footing.

\subsection{Leptogenesis}
An attractive possibility would be if the baryon asymmetry could be explained in the Natural Neutrinos framework from new processes at the weak scale.  The models may realize the Sahkarov conditions \cite{Sakharov:1967dj}.  There is explicit lepton-number violation which feeds into baryon number violation while electroweak sphalerons are active.  It may be possible to achieve sufficient CP-violation though the complex phases in the Yukawa couplings and/or Majorana mass matrices.  The small Yukawa couplings and Majorana masses required for small neutrino masses may also lead to out-of-equilibrium processes.  In fact leptogenesis has been previously found to be possible in the Inverse and Linear seesaw models \cite{Blanchet:2010kw,Gu:2010xc}, giving support to the possibility that it may be possible in the Natural Neutrinos framework.

There is, however, a potential obstacle.  The coupling required by naturalness $\mathcal{L} \supset h h^\dagger N_c N/f$ may keep all of the RHN fields in thermal equilibrium in the early Universe and this may make it difficult to satisfy one of the Sahkarov conditions.  Achieving leptogenesis may then require appealing to additional out-of-equilibrium dynamics, such as resonant processes involving the small mass-splittings between the pseudo-Dirac RHN.  It would be interesting to understand in quantitative detail whether the Natural Neutrino models may explain the baryon asymmetry of the Universe.

\subsection{Dark Matter}
Can some of the top partner fields $N_i$, $N_i^c$ be dark matter candidates? Naively, it would seem that there are no new stable particles since any would-be $Z_2$ symmetry protecting the RHN is violated by the Yukawa couplings, allowing the RHN to decay.  However, this is not necessarily the case. 

For instance, suppose we make $N_3$,  $N_3^c$ odd under $Z_2$, while all other fields are even. Clearly, the top Yukawa coupling in Eq.~(\ref{eq:simplified}) respects this $Z_2$ and the Higgs mass protection works as before. The $Z_2$ symmetry will zero out entries with only one $N_3$ or $N_3^c$ field in the neutrino mass matrix (\ref{eq:mass-matrix}). However, there is in fact still enough freedom to generate three light neutrino masses. This can be seen by inspecting the determinant in Eq.~(\ref{eq:big-det}). The expression is written as a product of three determinants of $3\times 3$ matrices, and one can see that the first two determinants are clearly non-zero due to the presence of $M_N$. The final determinant in Eq.~(\ref{eq:big-det}) is also non-zero provided the remaining entries in both $M_D$ and $M_D^c$ are non-zero. Therefore, in this example, $N_3$, $N_3^c$ are stable and potential dark matter candidates, while simultaneously three light neutrino masses are generated through their couplings to the other RHN top partner fields. As mentioned in the introduction, Ref.~\cite{Poland:2008ev} previously demonstrated that neutral fermionic top partners are potential dark matter candidates, hence this may be an interesting future direction of study for the Natural Neutrinos framework.

\subsection{Majorana Top Partners}
One potential area for future development would be to realize Majorana RHN as top partners, rather than the vector-like RHN arising in the models studied here.  In principle it would seem possible to cancel quadratic divergences with a term $\mathcal{L} \supset h h^\dagger N^2/f$.  However in order to enforce the required coupling at the TeV scale it would be necessary to embed the top quarks (which are in a complex representation) and the Majorana RHN (in a real representation) into an incomplete multiplet of some UV symmetry.  We did not find such a symmetry structure, however it may be possible with further study.

\subsection{Connecting $N_c$ with $N_F$}
As the Natural Neutrinos scenario enforces equality between the number of flavors of RHN and the number of colors in QCD ($N_N = N_c$) it is tempting to speculate as to whether it might be possible to tie the number of SM fermion families to the number of RHN in some encompassing scenario, i.e $N_F=N_N$.  Due to the first equality such a construction would then realize the very attractive possibility that the resolution of the little hierarchy problem and the mechanism behind the generation of neutrino masses leads to the prediction that the number of fermion families must be equal to the number of colors in QCD, $N_F = N_c$.  However, we did not find any construction that achieves this goal and believe it may be difficult, especially as the Natural Neutrinos scenario treats the third generation of quarks separately from the first two.  

\section{Summary}

In summary, the `Natural Neutrinos' scenario proposed here represents a new class of bottom up `neutral naturalness' models which address the little hierarchy problem while explaining the absence of new colored fields at the LHC. In our framework, the neutral top partners are simultaneously the RHNs responsible for the generation of the light neutrino masses.  The models also enforce a novel connection between the number of vector-like RHN ($N_N$) and the number of colors in QCD ($N_c$).  These models may arise in a variety of pNGB Higgs scenarios and we have sketched three specific models to demonstrate this. The precise structure of neutrino mass generation is based on a collective breaking of lepton number, allowing for the possibility of large neutrino Yukawa couplings. The models predict a plethora of potential signals in low-energy tests of lepton universality and lepton flavor violation, as well as possible signatures at high energy collider experiments.

\section*{Acknowledgements}
The authors are grateful for conversations with Alex Azatov, Roberto Contino, Nathaniel Craig, Gian Giudice, Florian Goertz, Ben Gripaios, Tao Han, Roni Harnik, Andrey Katz, Simon Knapen, Giuliano Panico, Riccardo Rattazzi, Richard Ruiz, Jose Santiago, Florian Staub, Daniel Stolarski, and Pedro Schwaller. Both authors acknowledge support from a CERN COFUND Fellowship.

\appendix

\section{On Proton Decay}
In the model descriptions of \Sec{sec:models} the visible sector quarks are described as living within larger multiplets (specifically $\SU(6)$) that contain also the hidden sector quarks.  As these hidden sector quarks become the RHN, and lepton number is broken, it would seem then that the multiplet structure may lead to proton decay as any global baryon-number $\U(1)_B$ which acts on the full multiplets containing visible and hidden sector quarks together has clearly been broken.\footnote{We thank Ben Gripaios for raising this interesting point at the CERN-CKC Neutral Naturalness workshop.}  In this section we will demonstrate that proton decay may be avoided in the models presented in \Sec{sec:models}.

As a warm up let us consider the Twin Higgs model.  In this scenario the approximate Twin symmetry leads to the appearance of a global $\SU(6)\times \SU(4)$ symmetry, however in reality this is only apparent as there are two copies of the SM fields, and the SM quarks and Twin quarks do not actually live in $\SU(6)$ multiplets, but rather $\SU(3) \times \SU(3)$.  The appearance of a full $\SU(6)$ symmetry is a consequence of the exchange symmetry which identifies the gauge couplings of both $\SU(3)$ groups as being equal.  With proton decay in mind, the (anomalous) global $\U(1)$ symmetries are $\U(1)_B\times\U(1)_L\times\U(1)_{B^T}\times\U(1)_{L^T}$, where the latter two denote the Twin symmetries.  Thus in order to generate the neutrino masses as described the symmetry is broken to $\U(1)_B\times\U(1)_L\times\U(1)_{B^T}\times\U(1)_{L^T} \to \U(1)_B\times\U(1)_{L-B^T}\times\U(1)_{L^T}$ and Twin baryon number breaking does not imply any breaking of the SM $\U(1)_{B}$ baryon number and the proton does not decay.

The situation for the Twin Higgs is similar to the case for the other models as in pNGB scenarios typically only subgroups of a larger global symmetry group are gauged.  Thus on a case-by-case basis additional global baryon-number symmetries which arise as elements of the full symmetry group may typically be imposed which protect the proton from decay.  However in order to demonstrate the origin of proton stability within the context of the full symmetry group it will be useful to consider the required global symmetry structure and discuss the gauged subgroups at a later stage.  This is useful because once subgroups are gauged it is sufficient to demonstrate that this gauging does not break the relevant $\U(1)$ factors.  Thus demonstrating a conserved baryon-number in the context of the full global symmetry structure is a useful first step.

We will now consider another example with an explicit embedding of the visible and hidden quarks into the same multiplet.  The symmetry breaking pattern in this section is inspired by earlier models based on $\SU(3)_W$ symmetries, as described in \cite{Schmaltz:2002wx,Kaplan:2003uc,Contino:2003ve,Perelstein:2003wd,Schmaltz:2004de,Berger:2012ec}.  As described, we first consider all of the symmetries as global symmetries, as pNGB scenarios typically rely on gauging only some subgroup of the full global symmetry group this is in keeping with the standard structure of a pNGB scenario.  We will return to the gauged subgroups later.  We will also only consider the third generation fermions.

\begin{table}[h]
\caption{\label{tab:symm}  Symmetries of the full matter content.}
  \begin{tabular}{ c | c | c | c | c }
    Field & $\SU(6)$ & $\U(1)_B$ & $\SU(3)_W$ & $\U(1)_X$ \\ \hline
    $Q$ & $\mb{6}$ & $1$ &  $\mb{3}$ & $-1/3$ \\
        $U^c \times 2$ & $\overline{\mb{6}}$ & $-1$ & $\mb{1}$ & $0$ \\
            $D^c$ & $\overline{\mb{6}}$ & $-1$  & $\mb{1}$ & $1$ \\
                $L$ & $\mb{1}$ & $0$ & $\mb{3}$ & $-1/3$ \\
                    $E^c$ & $\mb{1}$ & $0$  & $\mb{1}$ & $1$ \\
                        $H_1$ & $\mb{1}$ & $0$ &  $\mb{3}$ & $-1/3$ \\
                            $H_2$ & $\mb{1}$ & $0$ & $\mb{3}$ & $-1/3$ \\
                            $\phi \times 3$ & $\mb{6}$ & $1$ &  $\mb{1}$ & $0$
  \end{tabular}
  \label{tab:symm}
\end{table}

In \Tab{tab:symm} we show the field content and full symmetry group of the relevant matter in a UV completion of the $\SU(3)_W$ model detailed in \Sec{sec:models}.   The fields $H_{1,2}$ and $\phi_{1,2,3}$ are scalars.  $H_{1,2}$ will eventually be responsible for breaking $\SU(3)_W \times \U(1)_X \to \SU(2)_W \times \U(1)_Y$, however this aspect will not be relevant for proton decay.  As we will see, on the other hand, $\phi_{1,2,3}$ will be responsible for breaking $\U(1)_B \times \SU(6) \to \U(1)_{B_V} \times \SU(3)$ where $\U(1)_{B_V}$ is baryon number for the visible sector quarks.

The SM couplings arise from the following terms
\be
\mathcal{L} \supset Q H_1^\dagger U_1^c +Q H_2^\dagger U_2^c + \frac{1}{\Lambda} Q H_1 H_2 D^c + \frac{1}{\Lambda} L H_1 H_2 E^c ~~,
\ee
and the neutrino masses from
\be
\mathcal{L} \supset  \frac{1}{\Lambda} L H_1^\dagger \phi U^c + \frac{1}{\Lambda^3} (Q \phi^\dagger H_1^\dagger)^2 ~~.
\ee
Let us first consider the breaking of the full color group.  We will assume a scalar potential at the cutoff where $\phi_{1,2,3}$ all obtain VEVs in the last three components such that they break $\SU(6)\to \SU(3)$.  We can determine the remaining $\U(1)$ symmetries by considering the full set of Abelian generators acting on the quarks.  We may choose an arbitrary basis for the $\SU(6)$ generators and we thus choose to consider the basis of generators described in \Tab{tab:u1}.
\begin{table}[h]
\caption{\label{tab:u1}  $\U(1)$ Generators.}
  \begin{tabular}{ c | c }
    Generator & Diagonal Elements \\ \hline
    $\lambda_1$ & $\frac{1}{\sqrt{2}}(0,0,0,1,-1,0)$ \\
        $\lambda_2$ & $\frac{1}{\sqrt{6}}(0,0,0,1,1,-2)$ \\
            $\lambda_3$ & $\frac{1}{\sqrt{6}}(1,1,1,-1,-1,-1)$ \\
                $\lambda_4$ & $\frac{1}{\sqrt{2}}(1,-1,0,0,0,0)$ \\
                    $\lambda_5$ & $\frac{1}{\sqrt{6}}(1,1,-2,0,0,0)$ \\
                            $\lambda_B$ & $q_B \frac{1}{\sqrt{6}}(1,1,1,1,1,1)$ \\
                                $\lambda_X$ & $q_X \frac{1}{\sqrt{6}}(1,1,1,1,1,1)$ \\
  \end{tabular}
  \label{tab:u1}
\end{table}

When $\langle \phi_{1,2,3} \rangle \neq 0$ the generators $\lambda_{1,2,3,B}$ all appear to be broken.  However only one diagonal combination of the final two are broken and the full set of broken generators is $\lambda_1, \lambda_2, (\lambda_{3}-\lambda_B)$.  The unbroken ones are $\lambda_4, \lambda_5, (\lambda_{3}+\lambda_B)$ and $\lambda_X$.  The generator $\lambda_X$ is unbroken as $\phi$ carries no charge under this symmetry.  The generators $\lambda_4, \lambda_5$ correspond to the full unbroken $\SU(3)$ symmetry which we identify as the color symmetry of the SM.  The final generator $ (\lambda_{3}+\lambda_B)$ is the most interesting.  By studying the quark charges under the parent symmetry $\U(1)_B \times \SU(6)$ shown in \Tab{tab:symm}, one finds that the remaining $\U(1)_{3+B} \equiv \U(1)_{B_V}$ symmetry is simply baryon number symmetry for the SM quarks.

Thus, to summarise, if in addition to the $\SU(6)$ symmetry we assume a global $\U(1)_B$ symmetry acting on all of the $\SU(6)$ quark multiplets in the usual way, then when the scalars $\phi_{1,2,3}$ obtain vevs to break $\SU(6)$ down to the SM color group an additional $\U(1)$ symmetry will remain unbroken.  This $\U(1)$ symmetry is simply the baryon number symmetry acting on the quarks of the SM and, as it is unbroken, this symmetry protects from proton decay as the proton is the lightest state carrying this baryon number.

\begin{table}[tt]
\caption{\label{tab:symmb}  Symmetries of the fields below the scale of $\SU(6)$ breaking.  All quarks are now written in lower case and the subscript $T$ denotes a hidden sector quark. Representations under the spontaneously broken $\SU(3)_T$ are also shown.}
  \begin{tabular}{ c | c| c | c | c | c  }
    Field & $\SU(3)_c$ & $\SU(3)_T$ & $\U(1)_{B_V}$ & $\SU(3)_W$ & $\U(1)_{Z}$ \\ \hline
    $q$ & $\mb{3}$ &$\mb{1}$ & $1$ &  $\mb{3}$ & $1/3$ \\
        $u^c \times 2$ & $\overline{\mb{3}}$ & $\mb{1}$ & $-1$ & $\mb{1}$ & $-2/3$ \\
            $d^c$ & $\overline{\mb{3}}$ & $\mb{1}$  & $-1$  & $\mb{1}$ & $1/3$ \\
                $q_T$ & $\mb{1}$ &$\mb{3}$ & $0$ &  $\mb{3}$ & $-1/3$ \\
        $u_T^c \times 2$ & $\mb{1}$ & $\overline{\mb{3}}$ & $0$ & $\mb{1}$ & $0$ \\
            $d_T^c$ & $\mb{1}$ & $\overline{\mb{3}}$ &  $0$  & $\mb{1}$ & $1$ \\
                $L$ & $\mb{1}$& $\mb{1}$ & $0$ & $\mb{3}$ & $-1/3$ \\
                    $E^c$ & $\mb{1}$& $\mb{1}$ & $0$  & $\mb{1}$ & $1$ \\
                        $H_1$ & $\mb{1}$& $\mb{1}$ & $0$ &  $\mb{3}$ & $-1/3$ \\
                            $H_2$ & $\mb{1}$& $\mb{1}$ & $0$ & $\mb{3}$ & $-1/3$
  \end{tabular}
  \label{tab:symmb}
\end{table}

We may continue to study this theory to understand how the electroweak symmetry of the SM emerges.  We will again choose to consider a rotated linear combination of the $\U(1)$ symmetries $\U(1)_{Z=X+2/3 B_V} \times \U(1)_{B_V} \equiv \U(1)_{X} \times \U(1)_{B_V}$.  Up to overall choices in normalization this has not broken any of the global symmetries, thus the original $\U(1)_{B_V}$ is still a symmetry of the theory, however this choice will be convenient for understanding the electroweak symmetry structure. At present the matter content and symmetry representations under the unbroken symmetries are given in \Tab{tab:symmb}.

The relevant interactions are now

\begin{eqnarray}
\mathcal{L} & \supset & q H_1^\dagger u_1^c +q H_1^\dagger u_2^c +q_T H_1^\dagger u_{T,1}^c +q_T H_1^\dagger u_{T,2}^c + \nonumber \\
& & \frac{1}{\Lambda} q H_1 H_2 d^c + \frac{1}{\Lambda} q_T H_1 H_2 d_T^c + \frac{1}{\Lambda} L H_1 H_2 E^c ~~
\end{eqnarray}
and the neutrino masses will arise from couplings
\be
\mathcal{L} \supset  \frac{v_\phi}{\Lambda} L H_1^\dagger u_T^c + \frac{v_\phi^2}{\Lambda^3} (q_T H_1^\dagger)^2 ~~.
\ee
The extended electroweak symmetry structure now mimics the set up found in more conventional composite Higgs models such as \cite{Schmaltz:2002wx,Kaplan:2003uc,Contino:2003ve,Perelstein:2003wd,Schmaltz:2004de,Berger:2012ec} and it is clear that a pNGB with uncolored top partners will arise in this scenario.  No fields which obtain a vev from this point onwards are charged under the $\U(1)_B$ global symmetry, thus it is clear that proton decay is avoided at the level of only global symmetries.  However, for the sake of completeness we will now describe how the electroweak gauge group emerges.

Both of $H_{1,2}$ obtain vevs and thus break $\SU(3)_W\times \U(1)_Z \to \SU(2)_W \times \U(1)_Y$, where the $\U(1)_Y$ symmetry is a diagonal combination of the $\SU(3)_W$ generator proportional to $(1,1,-2)$ and the $\U(1)_Z$ generator proportional to $(1,1,1)$.  The charges of the remaining fields under this final symmetry group is shown in \Tab{tab:symmc} and the Hypercharge is identified as $Y=Z+T_8$ where $T_8 = 1/6(-1,-1,2)$

\begin{table}[tt]
\caption{\label{tab:symmc}  Symmetries of the fields below the scale of $\SU(6)$ breaking and after $\SU(3)_W\times \U(1)_Z \to \SU(2)_W \times \U(1)_Y$ breaking.  Fields which may be projected out by higher dimension boundary conditions are shown in red.}
  \begin{tabular}{ c | c | c | c | c | c }
    Field & $\SU(3)_c$& $\SU(3)_T$ & $\U(1)_{B_V}$ & $\SU(2)_W$ & $\U(1)_{Y}$ \\ \hline
    $q$ & $\mb{3}$ & $\mb{1}$ & $1$ &  $\mb{2}$ & $1/6$ \\
        $u^c$ & $\overline{\mb{3}}$ & $\mb{1}$ & $-1$ & $\mb{1}$ & $-2/3$ \\
            $d^c$ & $\overline{\mb{3}}$ & $\mb{1}$ & $-1$  & $\mb{1}$ & $1/3$ \\
                                $N$ & $\mb{1}$ & $\mb{3}$ & $0$ &  $\mb{1}$ & $0$ \\
        $N^c$ & $\mb{1}$ & $\overline{\mb{3}}$ & $0$ & $\mb{1}$ & $0$ \\
                $L$ & $\mb{1}$ & $\mb{1}$ & $0$ & $\mb{2}$ & $-1/2$ \\
                    $E^c$ & $\mb{1}$ & $\mb{1}$ & $0$  & $\mb{1}$ & $1$ \\
                        $h_1$ & $\mb{1}$ & $\mb{1}$ & $0$ &  $\mb{2}$ & $-1/2$ \\
                            $h_2$ & $\mb{1}$ & $\mb{1}$ & $0$ & $\mb{2}$ & $-1/2$ \\
                                                    $S_1$ & $\mb{1}$ & $\mb{1}$ & $0$ &  $\mb{1}$ & $0$ \\
                            $S_2$ & $\mb{1}$ & $\mb{1}$ & $0$ & $\mb{1}$ & $0$ \\
                                                            $L_S$ & $\mb{1}$ & $\mb{1}$ & $0$ & $\mb{1}$ & $0$ \\
                                    $\color{red}{T_L}$ & $\mb{3}$ & $\mb{1}$ & $1$ &  $\mb{1}$ & $2/3$ \\
                $\color{red}{T_R}$ & $\overline{\mb{3}}$ & $\mb{1}$ & $-1$ & $\mb{1}$ & $-2/3$ \\
                                $\color{red}{q_T}$ & $\mb{1}$ & $\mb{3}$ & $0$ &  $\mb{2}$ & $-1/2$ \\
                                        $\color{red}{u_T^c}$ & $\mb{1}$ & $\overline{\mb{3}}$ & $0$ & $\mb{1}$ & $0$ \\
                                                    $\color{red}{d_T^c}$ & $\mb{1}$ & $\overline{\mb{3}}$ & $0$  & $\mb{1}$ & $1$ \\
  \end{tabular}
  \label{tab:symmc}
\end{table}
This demonstrates that the required global symmetry structure is compatible with the electroweak gauge group and a conserved global $\U(1)_B$ baryon-number symmetry in the IR.

\subsection{Gauging subgroups}
While it is clear that the desired global symmetry structure may emerge in the IR, in pNGB scenarios subgroups of the global symmetries must be gauged to realize the SM gauge symmetry.  With regard to the electroweak gauge symmetry only the $\SU(2)_W \subset \SU(3)_W$ will be gauged.  If the full $\SU(3)_W$ symmetry were gauged the Higgs boson would be eaten and could not emerge as a pNGB.  The $\U(1)_Y$ subgroup must also be gauged to realize SM hypercharge.  This symmetry arose as a linear combination of a diagonal $\SU(6)$ generator, the original $\U(1)_B$ symmetry, and the original $\U(1)_X$ symmetry.  Thus the realization of hypercharge is necessarily linked to the full $\SU(6)$ symmetry structure.  Thus we will only gauge the $\SU(3) \times \SU(3) \subset \SU(6)$ factors of the color sector, and set the gauge couplings equal by an exchange symmetry.  This leaves all of the required $\U(1)$ generators, including the $\lambda_3 \propto 1/\sqrt{6} (1,1,1,-1,-1,-1)$ component of the full $\SU(6)$ symmetry, untouched.  We may then gauge the linear combination of these symmetries and identify it with hypercharge.  $\U(1)_{B_V}$ also emerges as an unbroken global symmetry as a consequence of the full $\SU(6)$ symmetry.

Considering the matter content that remains in the UV, shown in \Tab{tab:symmc}, we see that with this matter content hypercharge is anomalous.  We will in fact start out with the `twisted multiplets' of \Sec{sec:models}.  Taking these split multiplets removes precisely those fields in  red in \Tab{tab:symmc} and will leave hypercharge anomaly-free.  Simply removing these components of the full symmetry multiplets represents a hard breaking of the global symmetry, however enough of the symmetry remains to preserve the cancellation of quadratic divergences.  A UV justification for the removal of these states may be found in extra-dimensional theories and projection by boundary conditions. Crucially, this symmetry breaking does not introduce any breaking of the $\U(1)_{B_V}$ which prevents proton decay, thus it is innocuous from this perspective.

To conclude, we see that a stable proton is consistent with the full global symmetry structure in the UV, and with the breaking required to achieve the desired IR spectrum.  Furthermore, the symmetry which stabilizes the proton is also consistent with the hard global symmetry breaking introduced whenever subgroups of the full global symmetry are gauged.

\subsection{Relating to GUTs}
Finally we wish to comment on the difference between the models discussed here and proton decay in Grand Unified Theories (GUTs).  Since the hidden sector quarks mix with the SM neutrinos the hidden sector quarks effectively become leptons, thus the situation described here and the situation for GUTs is very analogous.  The source of proton decay in GUT theories is neatly described in \cite{Ross:1985ai}.  In GUTs if quarks and leptons live in the same multiplets proton decay does not arise.  The reason for this is that an effective baryon number is carried by the GUT gauge bosons, however this is still conserved and this alone cannot convert e.g.  $qqq \to l+\overline{q} q$.  The proton decays through the combination of quarks and leptons living in the same multiplet \emph{and} the fact that quarks and antiquarks live in the same multiplet, i.e. in the $\mb{10}$ of an $\SU(5)$ multiplet or a $\mb{16}$ of $\SO(10)$.  It is this collective symmetry breaking that allow proton decay.  One way of seeing this is that exchange of a GUT gauge boson may mediate $q q \to \overline{q} l$.

Returning to the Natural Neutrinos scenario, this second condition is not met.  Thus in principle if we were to gauge the full $\SU(6)$ symmetry then there would exist heavy gauge bosons which (after mixing through the neutrino mass terms) do couple quarks and leptons.  Processes such as $q l \to q l$ would be mediated by such gauge bosons.  However these gauge bosons would not also mediate $q q \to \overline{q} l$ as quarks and anti-quarks live in separate multipets and a global $\U(1)_B$ symmetry is preserved.

\bibliography{NaturalNeutrinosref}

\end{document}